\documentclass{pasa}%

\usepackage{graphicx}

\title[\textcolor{black}{The remnant radio galaxy associated with NGC~1534}]{\textcolor{black}{The remnant radio galaxy associated with NGC~1534}}

\author[\textcolor{black}{S.~W.~Duchesne \& M.~Johnston-Hollitt}]{
S.~W.~Duchesne$^{1,2,3}$ and 
M.~Johnston-Hollitt$^{1,2}$
\\
\affil{$^{1}$International Centre for Radio Astronomy Research (ICRAR), Curtin University, Bentley, WA 6102, Australia}
\affil{$^{2}$Peripety Scientific Ltd., PO Box 11355 Manners Street, Wellington, 6142, New Zealand}
\affil{$^{3}$School of Chemical and Physical Sciences, Victoria University of Wellington, P.~O. Box 600, Wellington 6140, New Zealand}
}

\jid{PASA}
\doi{10.1017/pas.\the\year.xxx}
\jyear{\the\year}

\usepackage{aas_macros}
\usepackage{hyperref} 
\hypersetup{colorlinks,citecolor=blue,linkcolor=blue,urlcolor=blue}

\usepackage{natbib}

\usepackage{pdflscape}  % Landscape pages
\usepackage{graphicx}	% Including figure files
\usepackage{amsmath}	% Advanced maths commands
\usepackage{amssymb}	% Extra maths symbols
\usepackage{amsfonts}   % For a checkmark
\usepackage{siunitx}    % For units.
\usepackage{multirow}   % For row contents that span multiple rows in tables.
\usepackage{booktabs}
\usepackage{longtable}
\usepackage[normalem]{ulem}

\usepackage{subcaption,cleveref}
\hypersetup{draft}
\newcommand{\tabft}[1]{(#1)}
\newcounter{ft}
\newcommand{\ft}[1]{\noindent%
  ~\refstepcounter{ft}\tabft{\alph{ft}\label{#1}}}
   
\DeclareSIUnit\parsec{pc}
\DeclareSIUnit\lightyear{ly}
\DeclareSIUnit\year{yr}
\DeclareSIUnit\jansky{Jy}
\DeclareSIUnit\beam{beam}
\DeclareSIUnit\ev{eV}
\DeclareSIUnit\erg{erg}
\DeclareSIUnit\mag{mag}
\DeclareSIUnit\rmtf{rmtf}
\DeclareSIUnit\gauss{G}

\def\degr{\ensuremath{^\circ}}
\def\arcmin{\ensuremath{'}}
\def\arcsec{\ensuremath{''}}
\def\hour{\ensuremath{^\mathrm{h}}}
\def\min{\ensuremath{^\mathrm{m}}}

\newcommand\fs{\mbox{$.\!\!^{\mathrm s}$}}%
\newcommand\farcs{\hbox{$.\!\!^{\prime\prime}$}}

\begin{document}

\begin{frontmatter}
\maketitle

\begin{abstract}
We present new observations of the large-scale radio emission surrounding the lenticular galaxy NGC~1534 with the Australia Telescope Compact Array and Murchison Widefield Array. We find no significant compact emission from the nucleus of NGC~1534 to suggest an active core, and instead find low-power radio emission tracing its star-formation history with a radio-derived star-formation rate of $0.38\pm0.03$~M$_\odot$\,yr$^{-1}$. The spectral energy distribution of the extended emission is well-fit by a continuous injection model with an `off' component, consistent with dead radio galaxies. We find the spectral age of the emission to be 203~Myr, having been active for 44~Myr. Polarimetric analysis points to both a large-scale magneto-ionic Galactic foreground at $+33$~rad\,m$^{-2}$ and a component associated with the northern lobe of the radio emission at $-153$~rad\,m$^{-2}$. The magnetic field of the northern lobe shows an unusual circular pattern of unknown origin. While such remnant sources are rare, combined low- and high-frequency radio surveys with high surface-brightness sensitivities are expected to greatly increase their numbers in the coming decade, and combined with new optical and infrared surveys should provide a wealth of information on the hosts of the emission.
\end{abstract}

\end{frontmatter}

\section{INTRODUCTION }
The active galactic nucleus (AGN) of a radio galaxy has a typical lifetime on the order of $\sim 10^8$ \si{\year} \citep{cor86}. These finite lifetimes give rise \textcolor{black}{to an observationally} rare stage of a radio galaxy's life where the radio plasma forming the lobes may remain visible after the core has shut down and \textcolor{black}{the supply of freshly accelerated plasma provided to the lobes by the resultant jets has ceased}. As energy loss due to synchrotron radiation is proportional to \textcolor{black}{square of the electron energy}, the highest-energy electrons lose energy more quickly \textcolor{black}{\citep[see][]{pac70}}, and these remnant lobes are usually observed with steep spectral indices, $\alpha$ \footnote{The spectral index, $\alpha$, is defined via $S \propto  \nu^\alpha$, where $S$ is the flux density at frequency $\nu$.}, \textcolor{black}{above some time-dependent break frequency} \citep{pmd+07,mpm+11,diw+14,bgm+16}. Such dying sources have predominantly been found within the dense environment of galaxy clusters where it is thought the external pressure from the intra-cluster medium (ICM) is able to stall the dissipation of the lobes \citep{mpm+11}. However, examples of dead and dying radio sources outside of clusters have been found \citep[e.g.][]{diw+14,bgm+16} and such sources in under dense environments have the potential to reach sizes in excess of $>700$~\si{\kilo\parsec} becoming so-called giant radio galaxy (GRG; e.g., \citealt{shsb05}). A GRG with an active AGN located in the field is expected to have a low surface brightness, and a dying, fading source much more so. Before the advent of low-frequency radio interferometers such as the Murchison Widefield Array \citep[MWA;][]{tgb+13, bck+13} and the LOw-Frequency ARray \citep[LOFAR;][]{lofar}, such sources---except in rare cases---were rendered mostly undetectable \textcolor{black}{due to their steep spectra in the GHz regime} \citep[see e.g.][]{cor87,gfgp88}.\par

A second class of radio galaxy with---at present---low known numbers are dust-rich disk galaxies (lenticular and spiral) with large-scale radio lobes. \textcolor{black}{Radio-loud AGN typically reside within large elliptical galaxies, whereas the denser material within spiral and lenticular galaxies may impede jet propagation}, resulting in the sub-kpc--scale jets often seen in Seyfert AGN \citep[e.g][]{uws81,gao+06}. At present there are only 9 spiral \citep{lok98,hso+11,bvv+14,mod+15,sis+15,mmm+16} and 4 lenticular (\citealt{bb57,cpc65,egks78}, Johnston-Hollitt et al. submitted) galaxies hosting large-scale radio emission, and \textcolor{black}{the cause of} their lack of detection is not clear as their radio powers, while lower on average than equivalent size radio galaxies, are still high enough to be detected by most radio instruments (Johnston-Hollitt et al., submitted).\par

NGC~1534 is a lenticular galaxy at a redshift of $z=0.017816$ \citep{dpd+91} and is part of the HDC~269 and LDC~292 galaxy groups \citep{chm+07}. \citet{hje+15} serendipitously discovered remnant radio emission surrounding NGC~1534 with the MWA. The large field of view of the observation and the sensitivity to large-scale structure enabled its detection. The low-surface brightness sensitivity of the MWA is due to the large number of short baselines between the antenna tiles, with a minimum baseline length of 7.7~m and 689 baselines $< 60~\si{m}$. In Phase I, the MWA had a maximum baseline of 2873.3~m, which at the MWA operating frequencies results in arcmin-scale synthesized beams. Despite this resolution limitation, large-scale extended structures can be studied in great detail due to the instrument's large fractional bandwidth, with individual observations able to observe with instantaneous bandwidths of 30.72~MHz. \citet{hje+15} considered the remnant radio plasma most likely associated with NGC~1534, considering it to be ancient lobes from a past cycle of core activity. This conclusion was motivated by the steep spectral index found, $\alpha_{185}^{843} = -2.1$, along with the general agreement in alignment between NGC~1534 and the remnant emission.\par

In this paper we present follow-up observations of the source with the Australia Telescope Compact Array \citep[ATCA;][]{fbw92} in the 16 \si{\cm} (2.1 GHz) and 15~\si{\milli\m} (17 and 19 GHz) bands as well as a complementary low-frequency analysis with data from the GaLactic and Extragalactic All-sky MWA survey \citep[GLEAM;][]{wlb+15}. \par

This paper assumes a flat $\Lambda$CDM cosmology, with $H_0 = 67.7$ \si{\kilo\m\per\s\per\mega\parsec}, $\Omega_\mathrm{m} = 0.307$, and $\Omega_\Lambda = 1-\Omega_\mathrm{m}$ \citep{aaa+16}. At the redshift of NGC~1534, 1 arcmin corresponds to 22.4 \si{\kilo\parsec}. For the calculation of image rms noise, we use the Background and Noise Estimation tool, \textsc{bane} \footnote{\url{https://github.com/PaulHancock/Aegean/wiki/BANE}}, part of the \textsc{AegeanTools} software package \citep{hmg+12,hth18}.

\section{Data}

\begin{figure*}
\centering
\includegraphics[width=0.75\linewidth]{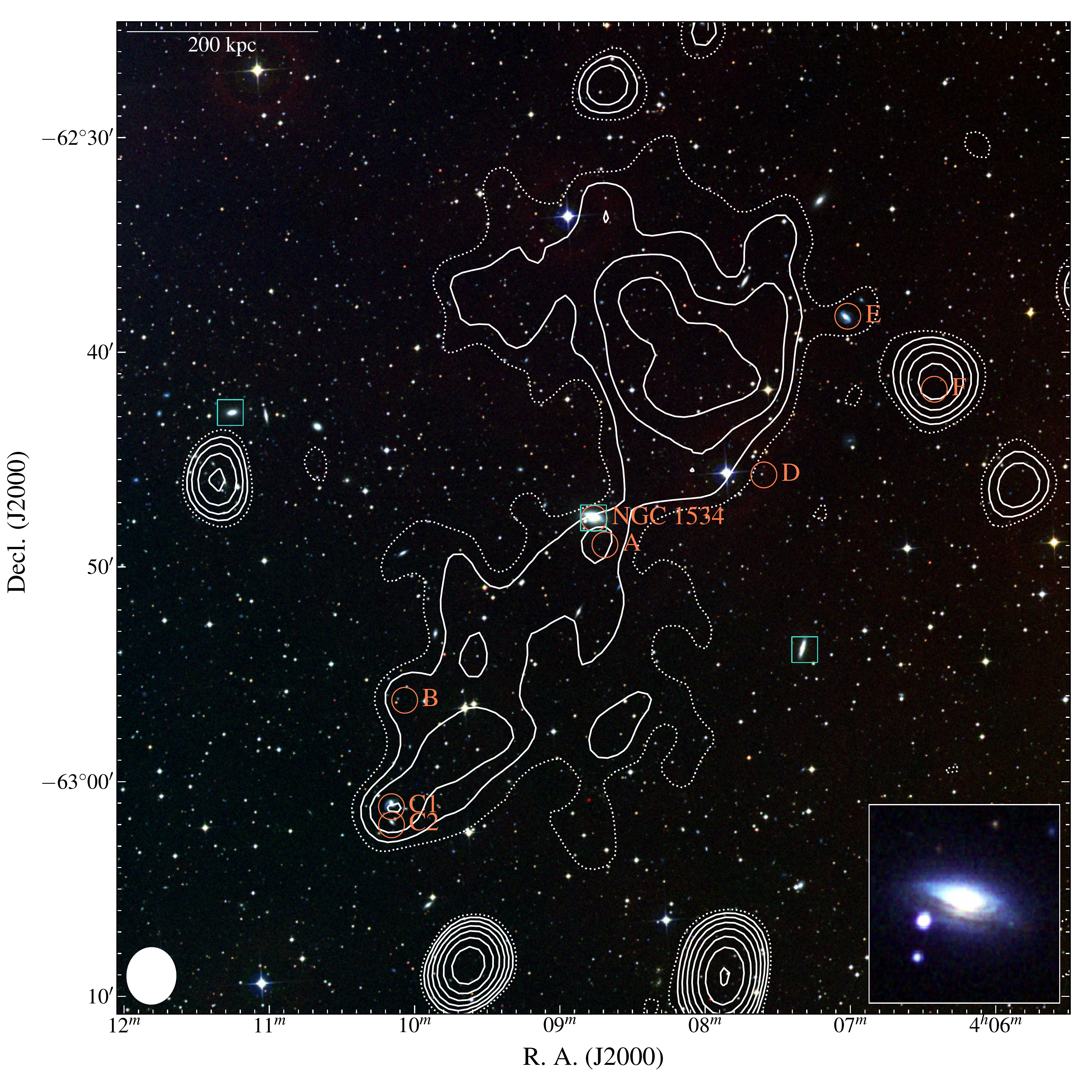}
\caption{The region surrounding NGC~1534. The background is an RGB image formed using the IR, red, and blue bands of the Digitized Sky Survey 2 (DSS2), and the contours overlaid are from the GLEAM survey at 200 \si{\mega\hertz}. Solid, white contours begin at 42.2 \si{\milli\jansky\per\beam} ($3\sigma_{\mathrm{rms}}$) increasing with $\sqrt{2}$. The dotted white contour is at $2\sigma_{\mathrm{rms}}$. The GLEAM data will be discussed in Section \ref{sec:mwa}. The linear scale is at the redshift of NGC~1534, and the inset is an RGB image formed using images generated by SuperCOSMOS \citep{supercosmos1,supercosmos2,supercosmos3}. Various sources are marked on the figure: orange circles are those that show significant radio emission that may add to low-resolution flux density measurements, discussed in Section \ref{sec:flux_density}; cyan squares are part of the group HDC~269, discussed in Section \ref{sec:host}. Other sources detected in the GLEAM image in this field are miscellaneous radio sources not discussed here. The white ellipse is the shape of the synthesised beam of the GLEAM data.}
\label{fig:NGC1534_rgb}
\end{figure*}

\subsection{MHz-frequencies---MWA}\label{sec:mwa}

\begin{table}
\caption{Properties of the wide GLEAM subbands. The central frequency, $\nu_{\mathrm{c}}$ is that specified in the FITS header of the survey products. The rms noise, $\sigma_{\mathrm{rms}}$, is calculated at the reference pixel of the FITS images, which are centred on NGC~1534.}\label{tab:gleam_subbands}
\centering
\resizebox{\linewidth}{!}{\begin{tabular}{c c c c}
\hline
\hline
Band & $\nu_{\mathrm{c}}$ & Beam shape & $\sigma_{\mathrm{rms}}$ \\
     & (\si{\mega\hertz})  & ($\arcmin \times \arcmin$, $\degr$) & (\si{\milli\jansky\per\beam}) \\
\hline
072-103 & 87.675 & $5.4 \times 5.0$, $-3.1$ & 49.1 \\ 
103-134 & 118.395 & $4.0 \times 3.6$, $2.8$ & 24.9 \\ 
139-170 & 154.235 & $3.1 \times 2.7$, $-5.4$ & 15.6 \\ 
170-231 & 200.315 & $2.6 \times 2.3$, $-1.8$ & 13.9 \\ 
\hline
\end{tabular}}
\end{table}
\setcounter{ft}{0}

The remnant emission surrounding NGC~1534 was originally detected serendipitously with the MWA at 185~\si{\mega\hertz} during a calibration pointing towards PKS~B0408-658 \citep{hje+15}. We confirm the detection of the emission with data spanning 72--231~MHz from the GLEAM survey. GLEAM is a recently completed survey of the southern sky ($\delta_{\mathrm{J2000}} < +25$) performed with the Phase I MWA using a drift scan imaging method to reduce primary beam calibration issues \citep{wlb+15}. The survey covers the frequency range 72--231~\si{\mega\hertz} with a declination- and band-dependent synthesized beam and sensitivities. One of the main products of the survey is the availability of three 30.72~\si{\mega\hertz}-bandwidth wideband images (hereafter 30-\si{\mega\hertz} subband images) as well as a single, more sensitive 60-\si{\mega\hertz} wideband image centred on 200.315~\si{\mega\hertz} (hereafter the 200-\si{\mega\hertz} band/image). Currently, imaging has been performed with a robust parameter of $-1$ in the `Briggs' weighting scheme \citep{bri95}---close to uniform weighting. The remnant radio emission is detected in the 200-\si{\mega\hertz} image. Fig. \ref{fig:NGC1534_rgb} shows the 200-\si{\mega\hertz} contours overlaid on the RGB optical image from the Digitised Sky Survey (DSS2). Additionally, the emission is detected in the three 30-\si{\mega\hertz} wideband images providing additional flux densities across the MWA band. GLEAM image properties are summarised in Table \ref{tab:gleam_subbands} and full imaging details can be found in \citet{gleamegc}.

\subsection{ATCA observations at 2.1 GHz}

\begin{table*}
\centering
\caption{Details for the ATCA observations. The scan time for mosaics is given as the total scan time for all three pointings. The frequency, $\nu$, is the observing frequency. A 2 \si{\giga\hertz} bandwidth is used for each observing frequency.}
\label{tab:atca_obs}
\centering
\resizebox{\linewidth}{!}{\begin{tabular}{c c c c c c}
\hline
\hline
Configuration & Date & $\nu$ & ${t_{\mathrm{scan}}}$ & Max. angular scale & Pointings \\
              &      & (\si{\giga\hertz}) & (\si{\minute}) & (arcmin) & ($\alpha_{\mathrm{J2000}}$, $\delta_{\mathrm{J2000}}$) \\
\hline
EW367         & 2014 Feb 25, 2014 Feb 26	   & 2.1 & 690 & \multirow{3}{*}{$ \left. {\begin{tabular}{c}13.0 \\ 19.6 \\ 18.7\end{tabular}} \right\}$} & \multirow{3}{*}{$ \left\{ \begin{tabular}{c}$04^\mathrm{h}07^\mathrm{m}47\fs875$, $-62\degr 36\arcmin 14\farcs 15$ \\ $04^\mathrm{h}08^\mathrm{m}46\fs099$, $-62\degr 47\arcmin 51\farcs 11$ \\ $04^\mathrm{h}09^\mathrm{m}45\fs091$, $-62\degr 59\arcmin 26\farcs 27$\end{tabular}\right.$}  \\
H75           & 2014 Apr 4     & 2.1 & 150 & & \\ 
H168          & 2017 Sept 28    & 2.1   & 270    & &  \\
H168          & 2016 Sept 3     & 17,19 & 315  & 1.1,1.2 & $04^\mathrm{h}08^\mathrm{m}46\fs070$, $-62\degr 47\arcmin 51\farcs 30$ \\
% H214          & 16-09-2014     & 5.5,9 & 10 & $\times$ \\ 
\hline
% \multicolumn{5}{l}{$^a$ For mosaic observations this is minimum time per pointing.}
\end{tabular}}
\end{table*}
\setcounter{ft}{0}

ATCA observations with the Compact Array Broadband Backend \citep[CABB;][]{cabb} \textcolor{black}{of the remnant emission and NGC~1534 were carried out at 2.1, 17, and 19~GHz}. Table \ref{tab:atca_obs} summarises the properties of the observations. The 2.1-\si{\giga\hertz} observations were carried out in 2014 and 2017 with the EW367 and H75 (PI Johnston-Hollitt), and H168 (PI Duchesne) configurations as part of project code CX287. PKS~B1934-638 is used as the bandpass, gains, and flux calibrator, and PKS~B0302-623 is used for phase calibration for the EW367 and H75 observations and PKS~0407-658 is used for the H168 observation. The use of PKS~0407-658 resulted in a loss of all antenna 6 data from this observation as the longer baselines show structure in the calibrator. Additionally, antenna 4 was unavailable for the 2.1-GHz H168 observations due to maintenance. Observations were performed as a 3-pointing mosaic to ensure the full extent of the emission was covered. \par

\begin{figure}
\includegraphics[width=0.99\linewidth]{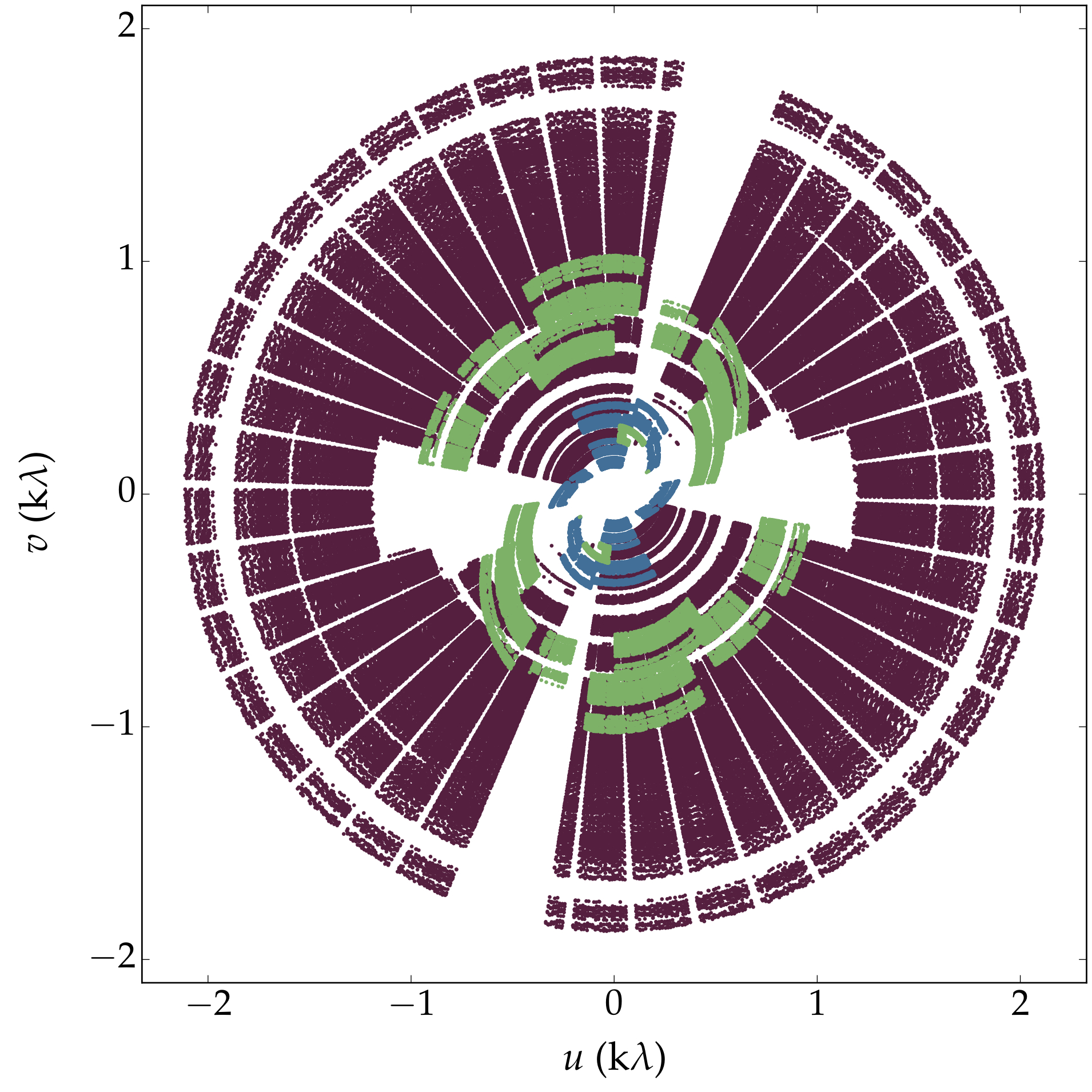}
\caption{The $u$--$v$ coverage for a single pointing of the combined EW367 (mauve), H75 (blue), and H168 (green) mosaic observations excluding antenna 6. Note that antenna 2 is missing from all H75 data, and antenna 4 is missing from all H168 data. This is for the 1510-\si{\mega\hertz} subband which features the most visibility flagging due to RFI.}
\label{fig:uvcover}
\end{figure}

\subsubsection{Calibration and flagging}

Data reduction follows standard reduction procedure using the software package \textsc{miriad} \citep{stw95}. In the following we briefly outline the process. The data are imported into \textsc{miriad} and bands with known RFI or self-generating interference are flagged, along with the 40 edge channels of the initial 2049 due to bandpass rolloff. The 2.1-\si{\giga\hertz} data are split into four subbands centred at 1510, 1942, 2375, and 2807 \si{\mega\hertz}, which are chosen to be 432 \si{\mega\hertz} to give equal frequency coverage based on the non-flagged channels. Calibration, and further RFI flagging, is performed for each of the subbands and pointings individually. We find that the lowest subband, at 1510 \si{\mega\hertz}, is more heavily affected by RFI reducing the usable data and resulting in a lessened sensitivity compared to the other bands. This is a common problem in the 1100--1400 \si{\mega\hertz} part of the 2.1-\si{\giga\hertz} band for the ATCA \footnote{\url{http://www.narrabri.atnf.csiro.au/observing/users_guide/html/atug.htmlInterference}} and has been noted by several authors \citep[e.g.][]{mfj+16,mjf+17,sjp16}. RFI flagging makes use of the \textsc{miriad} task \textsc{pgflag}, which utilises the \textsc{SumThreshold} method for detecting RFI in the $u$--$v$ data \citep{ovr12}. Calibration follows by first solving for complex gains and bandpass using the appropriate calibrator, then solving for complex gains and leakages on the secondary calibrators, finally applying a fluxscale correction based on PKS~B1934-638 and copying calibration solutions to the NGC~1534 pointings. After data are calibrated and flagged, we use a number of imaging processes to make subband and full-band images. Fig. \ref{fig:uvcover} shows the combined $u$--$v$ coverage for a single pointing (pointing 1) of the combined EW367, H75, and H168 data for the 1510-\si{\mega\hertz} subband after flagging. \par

% For the 17- and 19-\si{\giga\hertz} bands, we follow a similar process though due to the smaller fractional bandwidth we do not split these into subbands. 

\subsubsection{High-resolution imaging}\label{sec:highres}

\begin{figure*}
\centering
\includegraphics[width=0.8\linewidth]{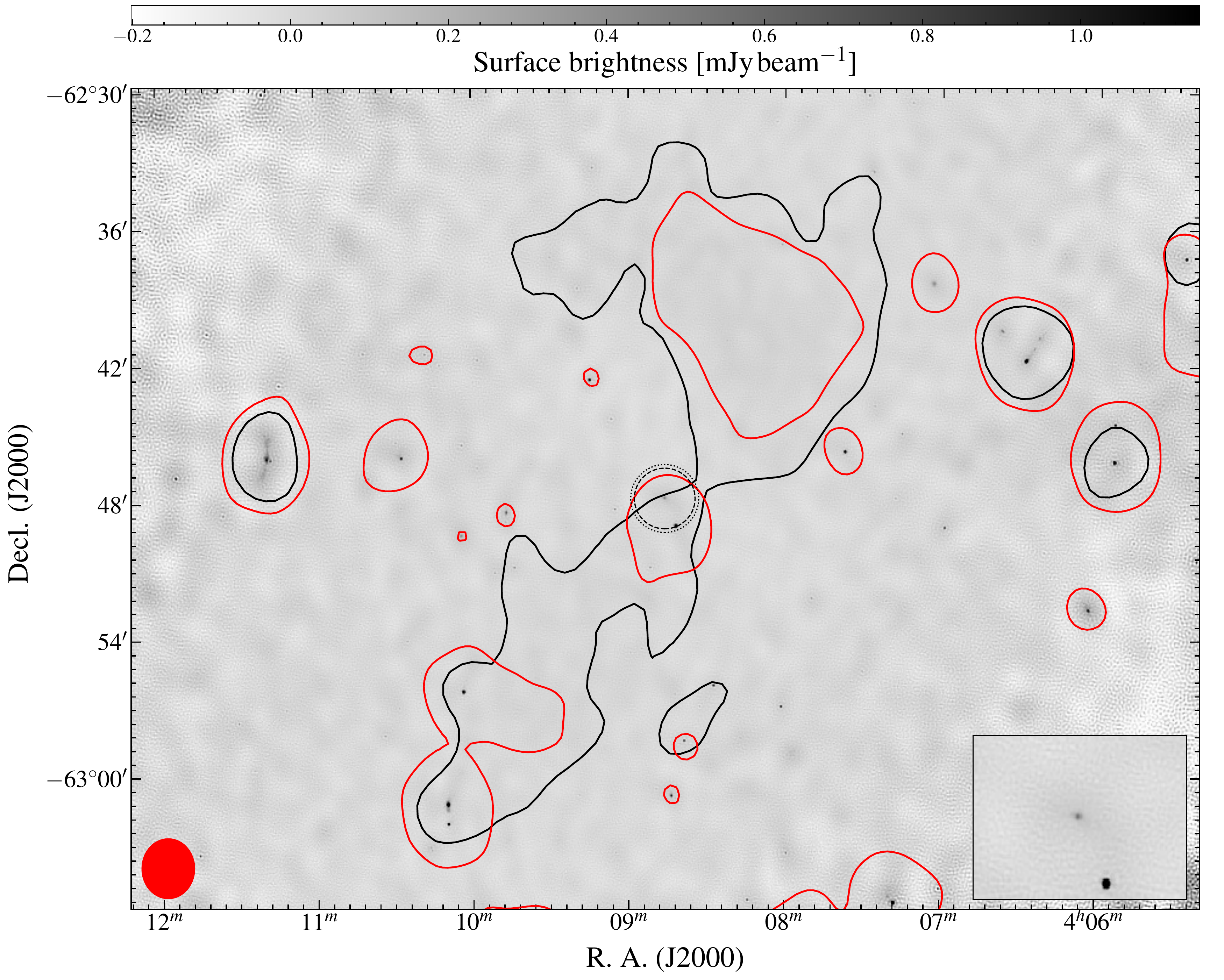}
\caption{High-resolution, stacked 2200-MHz ATCA image. The single, black contour is the GLEAM 200-MHz image at 43~\si{\milli\jansky\per\beam}, and the single, red contour is the low-resolution 1510-MHz ATCA image at 1.41~\si{\milli\jansky\per\beam}. The red ellipse in the lower-left is the beam shape of the low-resolution 1510-MHz image, and the black, dotted and dashed circles at the centre are the primary beams at 17 and 19 GHz, respectively. The inset shows this same central region.}
\label{fig:atca:highres}
\end{figure*}

\begin{figure*}
\centering
\includegraphics[width=0.8\linewidth]{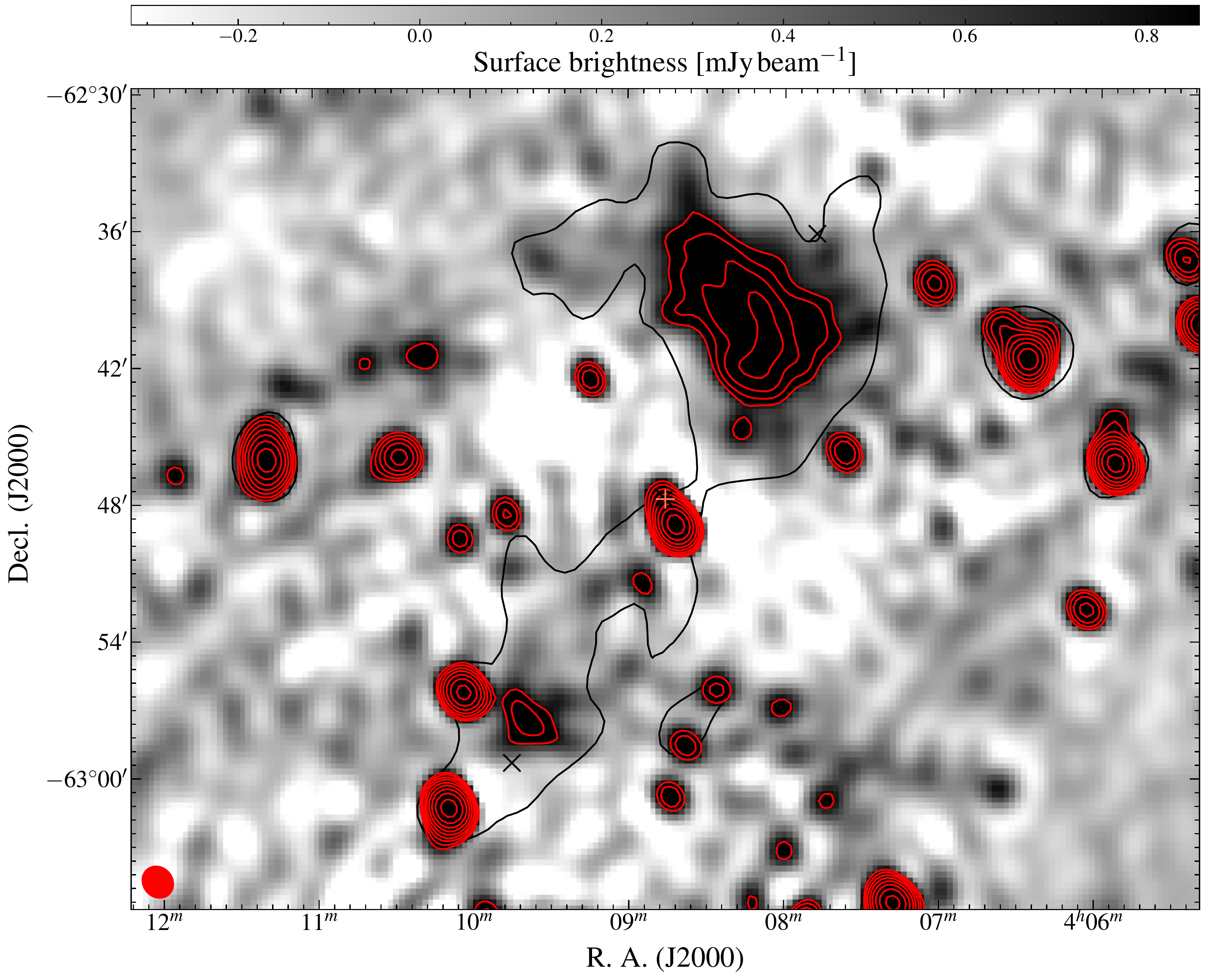}
\caption{Medium-resolution ($88\arcsec \times 73\arcsec$) 1510-MHz subband ATCA image. The single, black contour is as in Fig. \ref{fig:atca:highres}. The red contours are the 1510-MHz medium-resolution image, beginning at 810~\si{\micro\jansky\per\beam} and increasing with factors of $\sqrt{2}$. The red ellipse in the lower-left is the beam shape of the 1510-MHz image. The black \textcolor{black}{crosses} are the mosaic pointing centres, and the orange `$+$' indicates the position of NGC~1534.}
\label{fig:atca:medres}
\end{figure*}

\begin{table*}
\centering
\caption{ATCA image properties. The rms noise is the average at the centre of the map, as calculated by \textsc{bane}. For the 2200-MHz full-band image the higher-resolution, normally weighted images are convolved to a common beam shape (approximately equivalent to the 1510-MHz subband). The max angular scale is estimated from the minimum baseline of the H75 array (43~m without antenna 2) for the 2.1-GHz band images. Note that medium- and low-resolution images use a robust parameter of $+0.5$ whereas higher-resolution images use a robust parameter of $0$. Values in parenthesis are for (medium-resolution) and [low-resolution] images.}
\label{tab:atca_subband_properties}
\centering
\resizebox{\linewidth}{!}{\begin{tabular}{r c c c c c c}
\hline
\hline
Band & $\nu_{\mathrm{c}}$ & $\Delta\nu$ & Beam shape \tabft{\ref{tab:ref:taper1}} & $\sigma_{\mathrm{rms},I}$ & $\sigma_{\mathrm{rms},QU}$ \tabft{\ref{tab:ref:taper2}} & Max. angular scale \\
     & (\si{\mega\hertz}) & (\si{\mega\hertz}) & ($\arcsec \times \arcsec, \degr$) & (\si{\micro\jansky\per\beam}) & (\si{\micro\jansky\per\beam}) & (arcmin) \\
\hline
1510&1485.275& 432 &  $5.1 \times 4.6, \, 43$ &37(270)[470] & 100 & 19.4 \\ 
1942&1935.249& 432 & $4.3 \times 3.7, \, 59$ &34(150)[330] & 50 & 15.0\\ 
2375&2361.121& 432 & $3.6 \times 2.9, \, 67$ &46(120)[265] & 42 & 12.3\\ 
2807&2801.353& 432 & $3.3 \times 2.8, \, 62$ &41(110)[230] & 40 & 10.4\\ 
2200&2200.495& 1728 & $ 5.3  \times 4.6, \, 0$&   21& -& 13.3\\
17000&16852.728& 1849 & $0.72 \times 0.30, \, -0.7 \, \left( 13.2 \times 10.5, \, 74 \right) $ &18 (28) & - &1.2\\ 
19000&19090.607& 1849 & $0.64 \times 0.26, \, -0.6 \, \left( 11.8 \times 9.4, \, 74 \right) $ &20 (23) & -  &1.0\\
\hline
\end{tabular}}
\raggedright
\footnotesize{\emph{Notes.} \ft{tab:ref:taper1}~Medium- and low-resolution images have common beam shapes, unless otherwise specified: $(88\arcsec \times 78\arcsec)\left[157\arcsec \times136\arcsec\right]$. \ft{tab:ref:taper2}~For $88\arcsec \times 73\arcsec$ images.}
\end{table*}
\setcounter{ft}{0}

The first set of subband images we produced used the full set of visibilities and use a `Briggs' weighting scheme with robust parameter of 0 giving a balance between \textcolor{black}{resolution and sensitivity.} As the synthesized beam changes considerably across the bands we use the multi-frequency deconvolution task, \textsc{mfclean} \citep{mfclean}. After deconvolution, we perform one \textcolor{black}{cycle of phase-only self-calibration, CLEANing for more iterations in the second run of \textsc{mfclean}}. For the mosaic observation, this procedure is performed for each pointing, and the clean models, beam, and dirty maps are individually combined via the task \textsc{restor}. Finally the pointings are linearly mosaicked together with the task \textsc{linmos}. We also created a stacked full-band image by combining each pointing and subband image which maximises sensitivity which is shown in Fig. \ref{fig:atca:highres}. The properties of images produced are listed in Table \ref{tab:atca_subband_properties}. The remnant lobe emission is not detected though we find that NGC~1534 itself is detected across the 2.1-\si{\giga\hertz} band. 

% The bottom left image in Fig. \ref{fig:atca_examples} shows the 17-\si{\giga\hertz} robust 0 image with no emission detected from NGC~1534.

\subsubsection{Lower-resolution imaging}\label{sec:lowres}
\textcolor{black}{We also made two sets of images without antenna 6} (\textcolor{black}{hence removing baselines $> 367$~m}) to maximise sensitivity to large-scale structures. The procedure is the same as for the high-resolution images except we use a robust parameter of $+0.5$ to further increase sensitivity at a small cost to beam shape and do not phase calibrate, as the significant residual phase errors were only present on baselines involving antenna 6. We designate this first set as `medium-resolution' images and they have a common beam size of $88\arcsec \times 73\arcsec$. The second set of images follows the first, but were convolved with a Gaussian kernel to match the resolution of the 200-MHz GLEAM wideband from which we measure the flux density of the remnant emission. These images are designated as \textcolor{black}{`low resolution'} and they have a common beam size of $157\arcsec \times 136\arcsec$. The northern emission is well-detected in the 1510- and 1942-MHz low- and medium-resolution images, though approaches the $3\sigma_{\mathrm{rms}}$ detection limit in the 2375- and 2807-MHz images. Fig. \ref{fig:atca:medres} shows the 1510-MHz medium-resolution image with the northern emission visible. The emission from the southern lobe is also detected in the 1510-MHz medium- and low-resolution images. The image properties are listed in Table \ref{tab:atca_subband_properties}. \par

% The 17- and 19-\si{\giga\hertz} low-resolution imaging uses \textsc{mfclean} and is done over the entire 17- and 19-\si{\giga\hertz} bands as the change is synthesized beam is small enough that the change is accounted for. These images only cover the area shown by the dotted and dashed, black circles in Fig. \ref{fig:atca_examples}, with the only sources detected NGC~1534 and SUMSS~J040841-624908 (Source A). The low-resolution 17-\si{\giga\hertz} image is shown in the bottom-right panel of Fig. \ref{fig:atca_examples}. Source A is detected at the highest resolution in the 17-\si{\giga\hertz} image. In the 19-\si{\giga\hertz} band, Source A lies on the edge of the FWHM of the primary beam. NGC~1534 is only detected in the low-resolution 17-\si{\giga\hertz} image. \textcolor{black}{The lack of detection of NGC~1534 at the full resolution is suggestive of a lack of compact nuclear radio emission. Rather, the emission we see from NGC~1534 at 17~\si{\giga\hertz} and in the 2.1-\si{\giga\hertz} subband images is from star formation in the disk.} \par

\subsection{ATCA observations at 17 and 19 GHz}

\begin{figure*}
\centering
\begin{subfigure}{0.5\linewidth}
\includegraphics[width=1\linewidth]{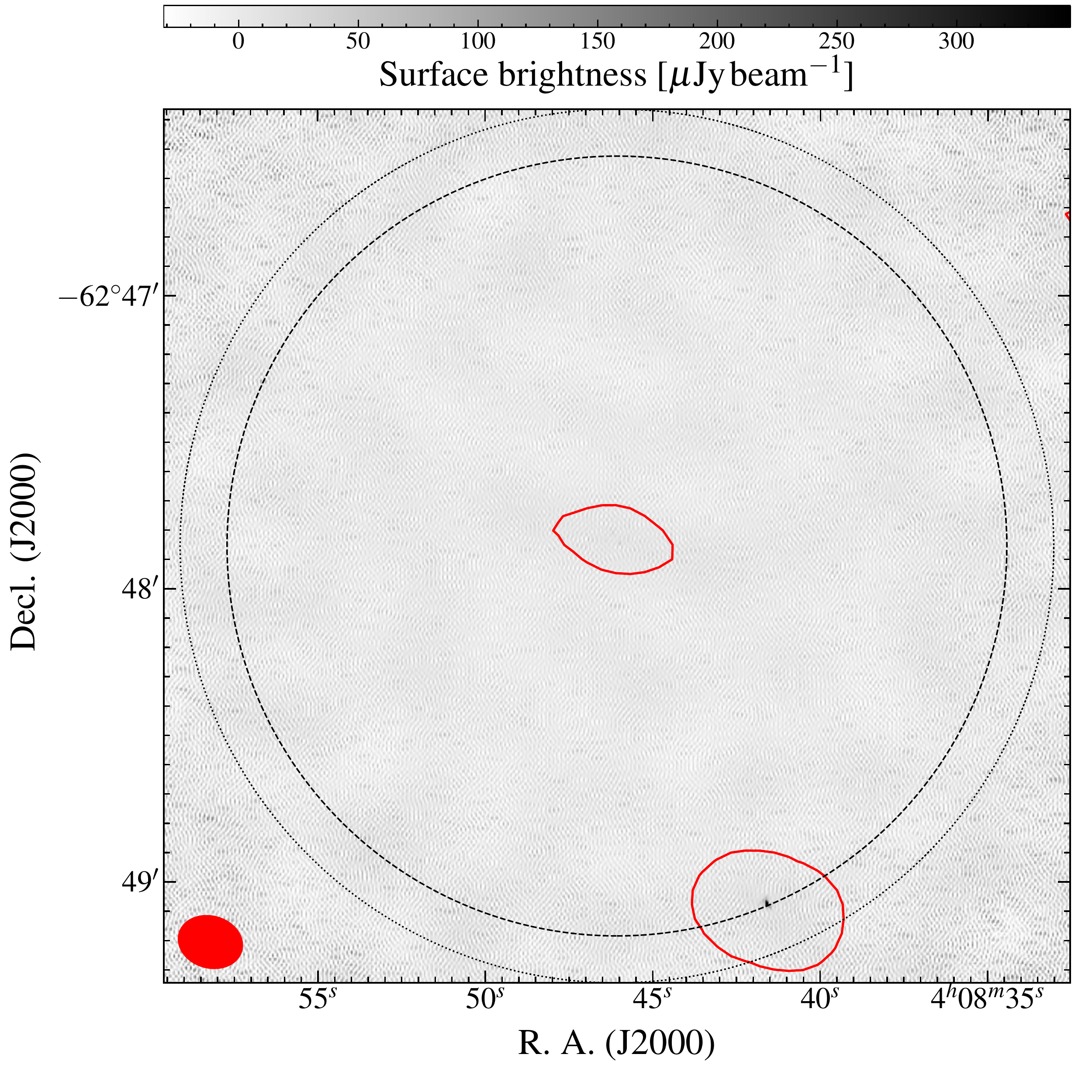}
\caption{\label{fig:15mm:highres}}
\end{subfigure}\hfill%
\begin{subfigure}{0.5\linewidth}
\includegraphics[width=1\linewidth]{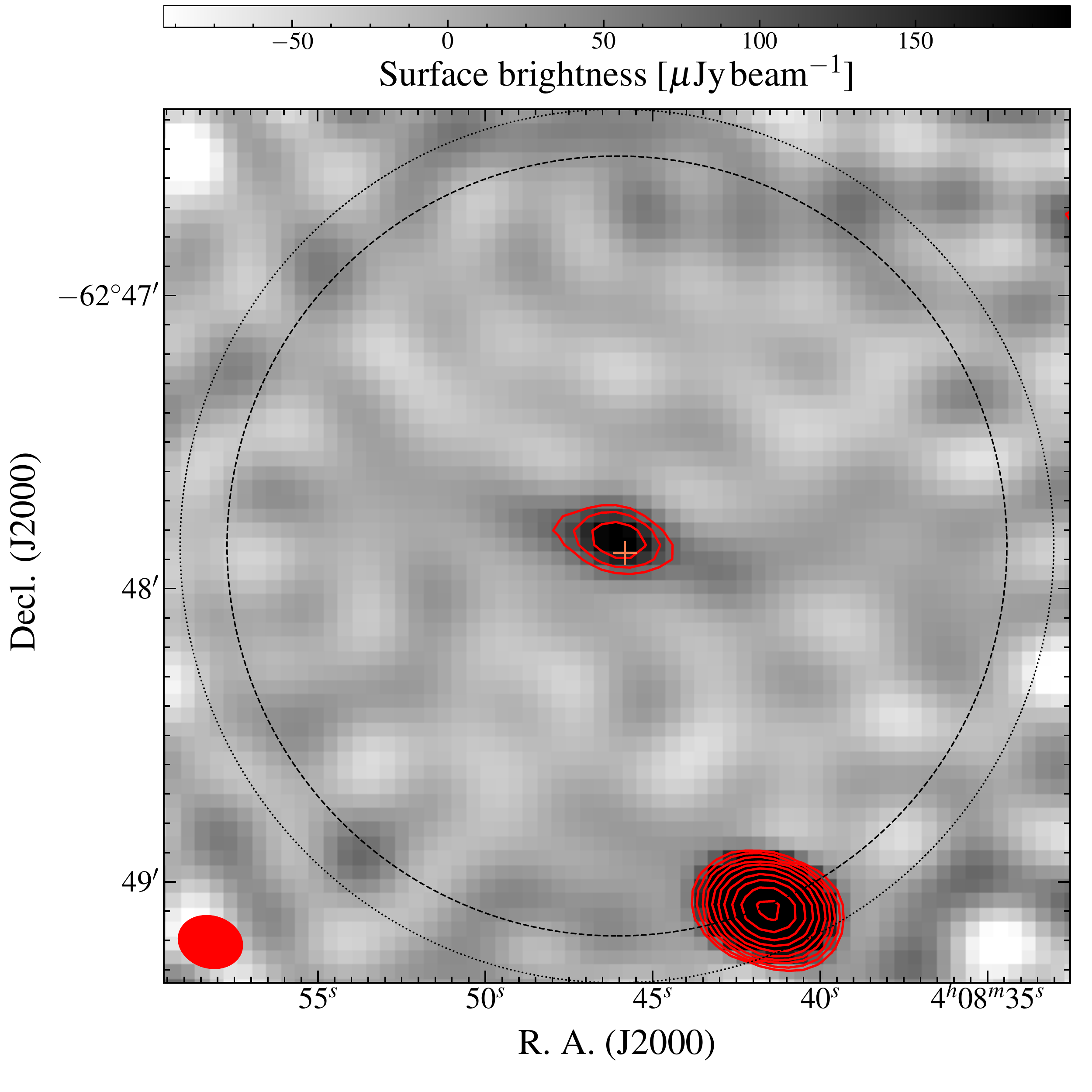}
\caption{\label{fig:15mm:medres}}
\end{subfigure}%
\caption{17 GHz ATCA images. \subref{fig:15mm:highres}: High-resolution, robust 0 image. \subref{fig:15mm:medres}: Medium-resolution, robust $+0.5$ image. The red contour(s) in both images are of the medium-resolution 17-\si{\giga\hertz} image starting at 84~\si{\micro\jansky\per\beam}. The dotted and dashed circles are the FWHM of the ATCA primary beam at 17 and 19~\si{\giga\hertz}, respectively, and the red ellipse in the lower-left corner is the beam shape of the 17-GHz medium-resolution image. The orange `$+$' in \subref{fig:15mm:medres} is the position of NGC~1534.}
\label{fig:atca_examples}
\end{figure*}

Complementary to the 2.1-\si{\giga\hertz} observations of the entire source, an observation at 17 and 19 \si{\giga\hertz} of the core of NGC~1534 was carried out with the H168 configuration in 2016 (Project Code CX366; PI Hurley-Walker). This was a single pointing, with NGC~1534 at the phase centre. The primary beam in this band is significantly smaller and only encompasses NGC~1534 and the nearby radio source, SUMSS~J040841-624908 (Source A in Fig. \ref{fig:NGC1534_rgb}). These observations were performed in an attempt to constrain the spectral index of NGC~1534, the putative host of the radio emission. This observation similarly used PKS~B1934-638 for flux calibration, though required PKS~B1921-293 for bandpass calibration and PKS~B0516-621 for both phase and pointing calibration. The observation details are presented in Table \ref{tab:atca_obs}. \par
\subsubsection{Calibration, flagging, and imaging}
The 17 and 19 GHz data reduction followed a similar procedure to the 2.1-GHz reduction, though RFI is less problematic and subbands are not made due to the smaller fractional bandwidth. As with the 2.1-GHz data, we make high-resolution, robust 0 images as well as medium-resolution, robust $+0.5$ images without antenna 6. The 17-GHz images are shown in Fig. \ref{fig:atca_examples}, and image properties are listed in Table \ref{tab:atca_subband_properties}. \textcolor{black}{Note} that there was no emission detected in the 19 GHz image and thus it is not considered any further.

\section{Analysis}
\subsection{Radio flux density}\label{sec:flux_density}
We measured the integrated radio flux density of the remnant emission from the GLEAM subbands and also estimated limits from the 2.1~\si{\giga\hertz} ATCA data. We begin by estimating contribution of flux density from interloping radio sources.

\subsubsection{Interloping radio sources}\label{sec:interlopers}

\begin{table*}
\caption{Spectral properties of sources marked in Fig. \ref{fig:NGC1534_rgb}. These sources, along with the core of NGC~1534, are the main sources of additional flux density GLEAM images. Integrated flux densities in the ATCA subbands are measured down to $3\sigma_{\mathrm{rms}}$, where $\sigma_{\mathrm{rms}}$ is computed for each pixel by \textsc{bane} except in the case of NGC~1534 where we measure down to $2\sigma_\mathrm{rms}$. The spectral index is calculated between the lowest- and highest-frequency measurements. Dashes in the flux density columns indicate no measurement available. 843-MHz measurements are made using \textsc{aegean}/\textsc{python} except for Sources C1/C2.}
\label{tab:misc_sources}
\centering
\resizebox{\linewidth}{!}{\begin{tabular}[width=\textwidth]{l l c c c c c c c}
\hline
\hline 
ID & Name & $S_{843}$ & $S_{1510}$ & $S_{1942}$ & $S_{2375}$ & $S_{2807}$ & $S_{17000}$ & $\alpha$ \\
   &      & (\si{\milli\jansky}) & (\si{\milli\jansky}) & (\si{\milli\jansky}) & (\si{\milli\jansky}) & (\si{\milli\jansky}) & (\si{\milli\jansky}) & \\
\hline
A & SUMSS~J040841-624908 &  $ 18.5\pm1.4$ & $12.99\pm0.27$ & $10.70\pm0.22$ & $8.89\pm0.18$ & $7.96\pm0.16$ &  $2.28\pm0.06$ & $-0.71\pm0.01$ \\ 
B & SUMSS~J041003-625615 &  $ 14.5\pm1.4$ & $8.46\pm0.18$ & $6.36\pm0.13$ & $5.36\pm0.12$ & $4.58\pm0.10$ & - & $-0.96\pm0.04$ \\ 
C1 & 2MASX~J04100936-6301152 & \multirow{2}*{$27.8\pm2.6$ \tabft{\ref{sumss_ref}}} &  $15.34\pm0.41$ & $12.99\pm0.32$ & $11.21\pm0.30$ & $10.06\pm0.25$ & - & $-0.67\pm0.05$ \\ 
C2 & 2MASX~J04100935-6302062 & &  $3.86\pm0.10$ & $3.06\pm0.08$ & $2.74\pm0.08$ & $2.40\pm0.07$ & - & $-0.73\pm0.06$ \\ 
D & GALEXASC J040736.82-624549.3 & - & $2.69\pm0.09$ & $3.68\pm0.12$ & $3.83\pm0.15$ & $4.22\pm0.15$ & - & $0.69\pm0.07$ \\  
E & PGC 014482 & - & $2.05\pm0.32$ & $1.77\pm0.20$ & $1.53\pm0.17$ & $0.99\pm0.14$ & - & - \\
F & SUMSS~J040627-624144 &$ 21.6\pm1.6$ \tabft{\ref{F_ref}} & $9.70\pm0.20$ & $7.21\pm0.16$ & $4.64\pm0.14$ & $2.69\pm0.15$ & - & - \\ 
- & NGC~1534 & - & $2.20\pm0.16$ & $1.39\pm0.09$ & - & $1.12\pm0.09$ & $0.301\pm0.048$ & $-0.79\pm0.08$ \\ 
\hline
\end{tabular}}
\footnotesize{\emph{Notes.} \ft{sumss_ref}~Total flux density of C1 and C2 from the SUMSS catalogue \citep{mmg+07}; not used in fitting the spectral index. \ft{F_ref}~Total flux density of Source F and nearby sources; not used in fitting.}
\end{table*}
\setcounter{ft}{0}

\begin{figure}
\includegraphics[width=0.99\linewidth]{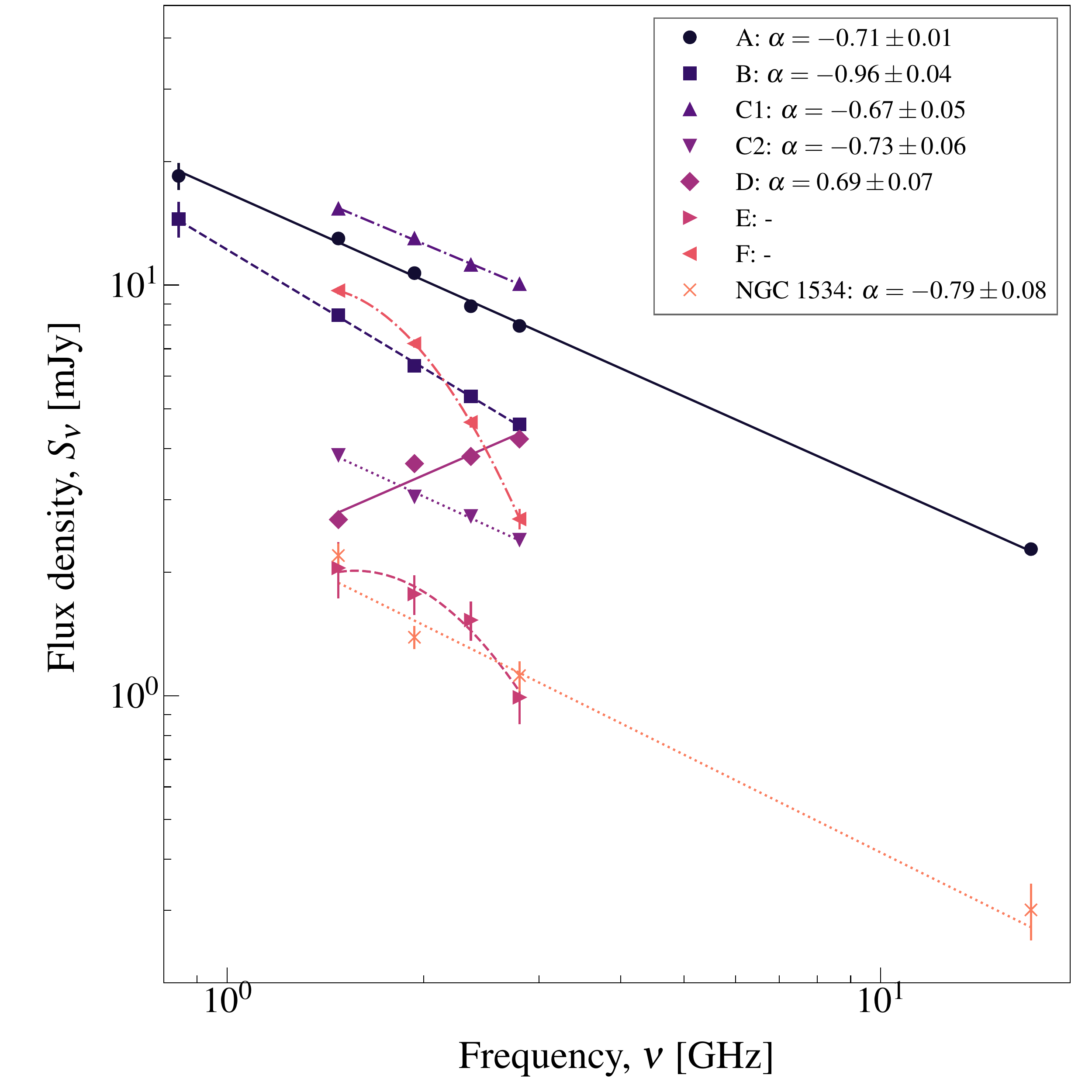}
\caption{The SEDs of sources within the remnant emission. The data are also presented in Table \ref{tab:misc_sources}. }
\label{fig:multi_sed}
\end{figure}

We measure the flux densities of the sources labelled in Fig. \ref{fig:NGC1534_rgb} across our ATCA subband images, as well as from 843-MHz data taken from the Sydney University Molonglo Sky Survey \citep[SUMSS;][]{bls99,mmb+03}. Table \ref{tab:misc_sources} summarises the flux density measurements, spectral indices, and gives the names of the sources. Fig. \ref{fig:multi_sed} plots the spectral energy distribution (SED) of each source. Source E is not a point source at the full resolution of the ATCA images and so we measure flux densities for this source from the low-resolution ATCA subbands, and Source F has extended emission to the north west in the high resolution ATCA images.
We use two methods for source measurements: for confirmed point sources, we utilise the source-finding, measuring, and characterising software, \textsc{aegean} \citep{hmg+12,hth18} with a detection threshold of $6\sigma_{\mathrm{rms}}$ and source growth threshold of $3\sigma_{\mathrm{rms}}$. Thus we are making sure sources are detected above $6\sigma_{\mathrm{rms}}$ and that they are being measured out to $3\sigma_{\mathrm{rms}}$. For other sources we use an in-house \textsc{python} code to identify connected pixels that comprise an extended source---\textcolor{black}{using a flood-fill algorithm as in \textsc{aegean} and measuring integrated flux density in the same manner as the source-finding software \textsc{duchamp} \citep{duchamp}. Error calculations are made using rms maps generated by BANE, allowing the rms to vary across the source, yielding \begin{equation}\label{eq:flux_unc}
\sigma_{S_\nu} = \sqrt{\left( f S_\nu\right)^2 + \left( \sum_i \sigma_{\mathrm{p},i}\right)^2} \quad [\mathrm{Jy}] \, ,
\end{equation}
where $\sigma_{\mathrm{p},i}$ is the rms at a specific pixel in Jy\,pixel$^{-1}$, and $f$ the additional uncertainty for the flux scale/calibration uncertainties of the specific map.}\par

Source A is a curious case as measuring the peak flux density and comparing to the integrated flux density shows a significant discrepancy in the SUMSS data. The integrated flux density is lower, at $S_{843} = 13 \pm 2$ \si{\milli\jansky} (cf. 843-\si{\mega\hertz} peak flux density measurement in Table \ref{tab:misc_sources}). The discrepancy is likely due the source's location within a negative bowl resulting in an underestimated flux density measurement. For consistency, we measure the peak flux density values of Source A for all measurements, and note that in all images Source A is unresolved. Further, Source A is at the edge of the 19-\si{\giga\hertz} primary beam, thus we do not measure the flux density in this band. For the emission from NGC~1534, the full resolution ATCA images show little nuclear activity, but detects extended emission in the disk of the galaxy, likely from star formation. We note the lower sensitivity of the 2375-MHz image made measurement of the the NGC~1534 emission problematic and no measurement there is provided. Most of the sources show typical powerlaw spectra of radio galaxies, with SEDs fit by \begin{equation}\label{eq:powerlaw}
S_\nu =  C \nu^{\alpha} \, ,
\end{equation} where $\alpha$ is the spectral index and $C$ the flux normalisation. For Sources E and F, the SEDs show significant curvature and are fit by a generic curved powerlaw model of the form \begin{equation}
S_\nu = C \nu^{\alpha} \mathrm{e}^{q \left(\ln \nu\right)^2} \, ,
\end{equation}
where $\alpha$ is the equivalent spectral index in the case of no curvature, and $q$ is the curvature index \citep[e.g.][]{db12,ceg+17}. In Table \ref{tab:misc_sources} we only report the power law index when using \textcolor{black}{Eq.} \ref{eq:powerlaw}.\par
Fitting is done via non-linear weighted least squares methods using the Lavenberg-Marquart algorithm implemented in \textsc{lmfit} \citep{lmfit}. The errors on the flux density measurements are the quadrature sum of the \textsc{aegean}/\textsc{python} measurements with the percentage uncertainty associated with the maps \textcolor{black}{(as in Eq. \ref{eq:flux_unc} for the in-house \textsc{python} code)}. For the ATCA, this is 2 \%~\citep[see e.g.][]{vbmh00, jsg+08}, and for the SUMSS map this is 3 \%~\citep{mmb+03}. \par
The power law model fit for NGC~1534 suggests a 1.4-\si{\giga\hertz} flux density of $1.97\pm0.15$. This translates to a 1.4-\si{\giga\hertz} power of $P_{1.4} = (1.5 \pm 0.1) \times 10^{21}$ \si{\watt\per\hertz}.

\subsubsection{The remnant radio emission}

\begin{table}
\caption{Flux density measurements of the total, northern, and southern lobe emission. Flux densities are measured out to $2\sigma_\mathrm{rms}$ as per \citet{hje+15}.}
\label{tab:relic_emission}
\centering
\resizebox{\linewidth}{!}{\begin{tabular}[width=\textwidth]{c c c c c}
\hline
\hline 
Band & ${S_\nu}^{\mathrm{north}}$ & ${S_\nu}^{\mathrm{south}}$ & ${S_\nu}^{\mathrm{total}}$ & Reference \\
(MHz) & (mJy) & (mJy) & (mJy) & \\
\hline

88  & $2270\pm220$ & $840\pm400$ & $2110\pm330$ &\tabft{\ref{this_work}} \\
118 & $1910\pm180$ & $740\pm320$ & $2650\pm260$ &\tabft{\ref{this_work}} \\
154 & $1670\pm150$ & $720\pm270$ & $2390\pm230$ &\tabft{\ref{this_work}} \\
408 & $520\pm104$ & $300\pm60$ & $820\pm120$ &\tabft{\ref{hje+15_ref}} \\
843 & $130\pm20$ & $80\pm20$ & $210\pm30$ &\tabft{\ref{hje+15_ref}} \\
1400 & $<45$ & $<27$ & $<72$ &\tabft{\ref{hje+15_ref}} \\
1510 & $35\pm1$ & $>10.2$ & $>45 \pm 1$ & \tabft{\ref{this_work}} \\
1942 & $13.4\pm0.7$ & - & $>13.4 \pm 0.7$ & \tabft{\ref{this_work}} \\
2375 & $>2.8$ & - & $>2.8$ & \tabft{\ref{this_work}} \\
2807 & $>1.9$ & - & $>1.9$ & \tabft{\ref{this_work}} \\
\hline
\end{tabular}}
\raggedright
\footnotesize{\emph{References.} \ft{this_work}~This work; \ft{hje+15_ref}~\citet{hje+15}.}
\end{table}

\setcounter{ft}{0}

We measure the integrated flux densities of the northern lobe from the GLEAM wideband images as well as the 1510- and 1942-MHz low-resolution ATCA images. Due to the blended nature of compact and extended emission within the southern lobe at MWA frequencies, we measure the integrated flux densities of the total emission in the GLEAM images, subtracting the northern lobe contribution for the initial estimate of the flux density of the southern lobe. As the southern lobe is only well-detected, and not blended in the 1510-MHz medium-resolution image, we measure it there. 2375- and 2807-MHz lower limits are placed on the northern lobe based on vague detection at $2\sigma_\mathrm{rms}$. similarly, a lower limit at 1942-MHz is placed on the southern lobe, though 2375- and 2807-MHz upper limits are not estimated here due to confusion with compact sources. \par
Flux densities for the extended emission \textcolor{black}{are measured using the in-house \textsc{python} code}, where we limit measured pixels to those above $2\sigma_{\mathrm{rms}}$ This $\sigma_\mathrm{rms}$ cut is chosen for consistency with \citet{hje+15} and because we have prior knowledge that the emission is of particularly low surface brightness. We consider the rms noise on a pixel-by-pixel basis using \textsc{bane}. \textsc{bane} uses sparse pixel grids to account for instances where noise may change rapidly across the image. Uncertainties in flux density measurements are given by Eq. \ref{eq:flux_unc}. We use the model parameters of the interloping sources to extrapolate to MWA frequencies for subtraction from GLEAM images, where appropriate. This is not necessary for the ATCA subband images as no significant interloping sources are found within the emission region at these frequencies. Table \ref{tab:relic_emission} summarises the measured flux densities, with additional literature data measured by \citet{hje+15} from SUMSS \citep{bls99,mmb+03}, a re-processed Molonglo Reference Catalogue image \citep{mrc}, and an upper limit from CHIPASS \footnote{\url{http://www.atnf.csiro.au/research/CHIPASS/}} \citep{chipass}. \par
Using the 200-MHz GLEAM image and the 1510-MHz low-resolution ATCA image, we estimate the projected size of the emission assuming it is indeed emission associated with NGC~1534. The projection separation between the peaks in the northern and southern lobes is $\sim 20$~arcmin which translates to a projected linear size of $\sim 450$~kpc at the redshift of NGC~1534. This is smaller than the size found by \citet{hje+15} though their estimate includes Source C1/C2 and extends further north. We do not include Source C1/C2 as there is no evidence that the emission continues beyond the southern peak at 1510 MHz. However, the emission may continue further northwest, in which case the projected size may be up to $\sim 610$~kpc. We cannot be sure this is the case, as there are a number of faint point sources which may be contributing to the morphology of the emission at the northwestern end. \par
%The ratio of brightness of the northern lobe to either NGC~1534 or Source A is suggestive of an FR-II radio source \citep{fr74}. If Source A is the host, the southern lobe may not follow this description, which would imply a hybrid morphology radio source \citep[HyMoRS; e.g.][]{deg17}, however, given the lower limit in the 1510-MHz flux density measurement and the likelihood of NGC~1534 being the host, the FR-II description is more appropriate.

\subsubsection{The spectral energy distribution}

\begin{figure*}
\begin{subfigure}{0.355\linewidth}
\includegraphics[width=1\linewidth]{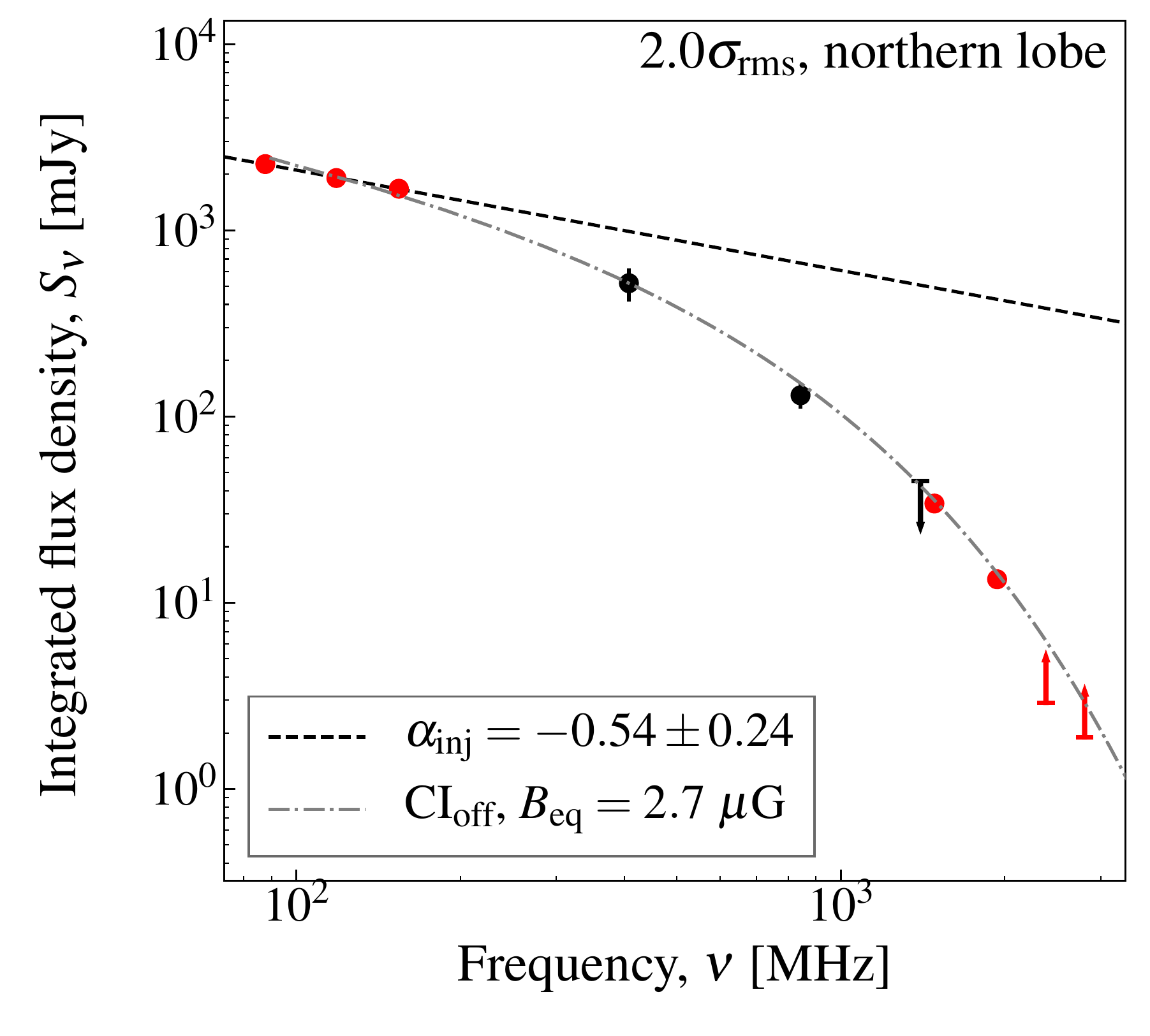}
\caption{\label{fig:seds:north}}
\end{subfigure}\hfill%
\begin{subfigure}{0.315\linewidth}
\includegraphics[width=1\linewidth]{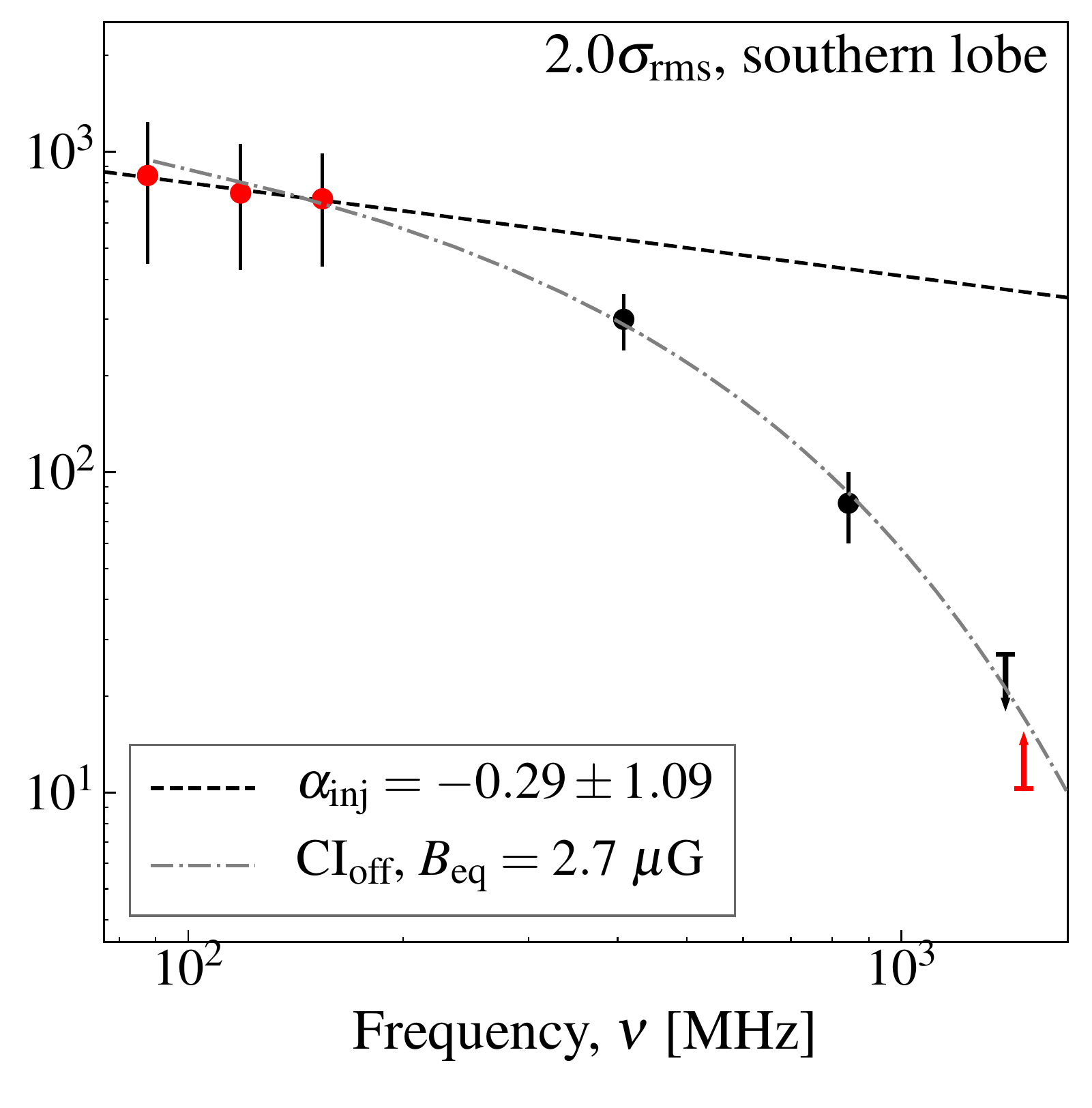}
\caption{\label{fig:seds:south}}
\end{subfigure}\hfill%
\begin{subfigure}{0.315\linewidth}
\includegraphics[width=1\linewidth]{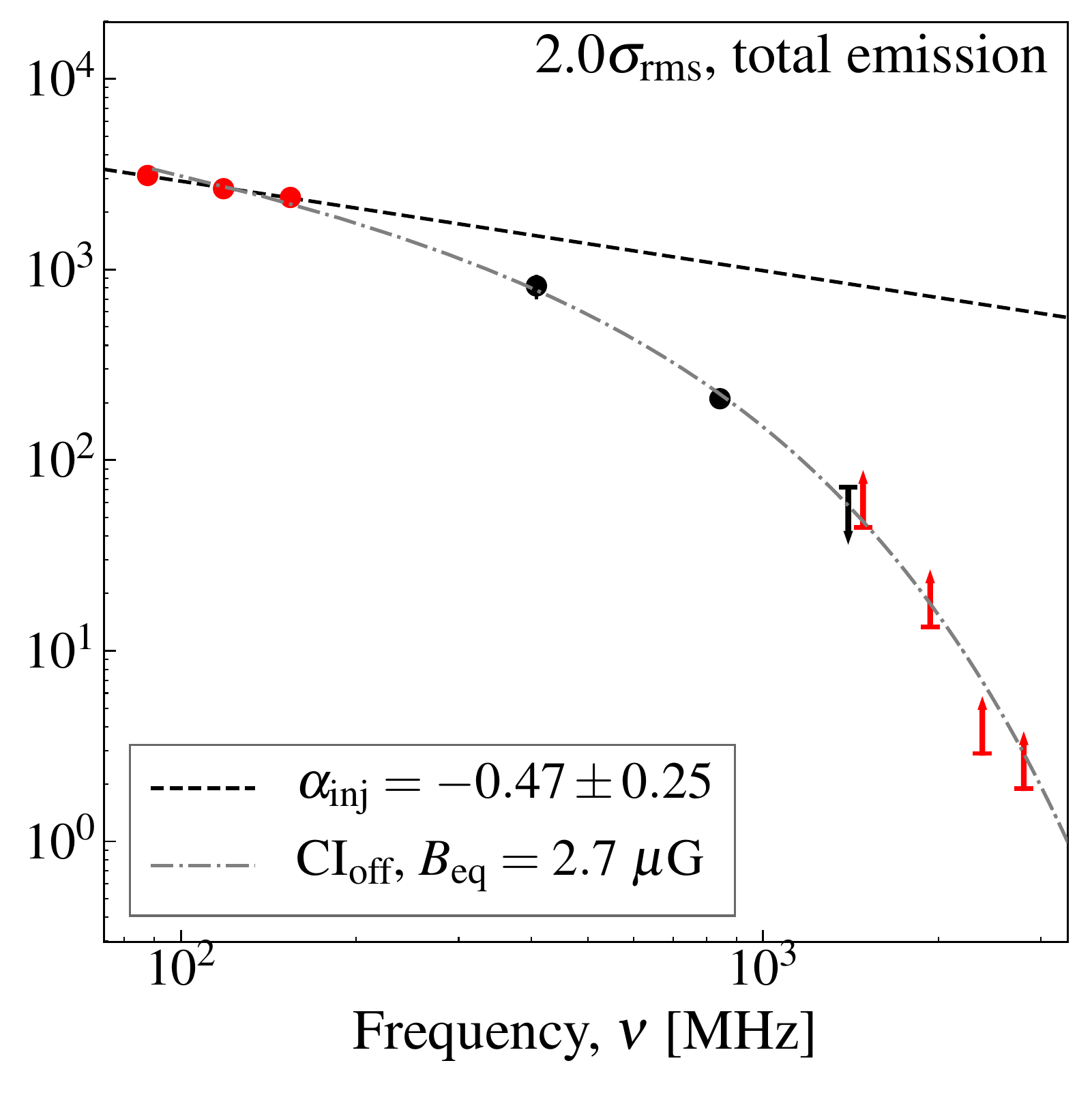}
\caption{\label{fig:seds:total}}
\end{subfigure}%
\caption{The SED of the emission surrounding NGC~1534 after a $2\sigma_{\mathrm{rms}}$ cut to the pixels. \subref{fig:seds:north}: emission from the northern lobe. \subref{fig:seds:south}: emission from the southern lobe. \subref{fig:seds:total}: combined emission from the northern and southern lobes. Measured flux densities have sources subtracted, where appropriate, based on spectral indices derived in Section \ref{sec:interlopers}. A $\mathrm{CI}_\mathrm{off}$ model is fit for the northern and southern lobes separately, then for the combined emission. Limits are indicated by arrows, points in black are from the literature (see Table \ref{tab:relic_emission}), and points in red are measured in this work. Limits are not used in the fitting process.}
\label{fig:seds}
\end{figure*}

In the frequency regime measured here, the SED is not described by a simple powerlaw model, and instead we consider the continuous injection (CI) models \citep{kar72,pac70,jp73} implemented in the Broadband Radio Astronomy ToolS  \citep[BRATS;][]{brats1,brats2} package \footnote{\url{http://www.askanastronomer.co.uk/brats/}}. The standard CI model is fit under the assumption the magnetic field is in equiparition with the emitting electron population. We assume that the AGN has switched off, as the ATCA data suggest no prominent nuclear activity---hence, we fit the $\mathrm{CI}_\mathrm{off}$ model which describes remnant radio emission described in \citet{kg94} as a modification to the CI model as described in \citet{jp73}. For $\mathrm{CI}_\mathrm{off}$ fitting we assume the emission is at the redshift of NGC~1534. \par
In fitting we require an injection index, \textcolor{black}{$\alpha_\mathrm{inj}=\left(1-\delta_\mathrm{inj} \right)/2$, that describes the observed emission from a continuous injection of fresh electrons with a power law energy distribution of index $\delta_\mathrm{inj}$, assuming synchrotron and inverse-Compton losses. Fig. \ref{fig:seds} shows power law fits to the GLEAM subband data from which we obtain $\alpha_\mathrm{inj}$, which is valid if a break frequency, $\nu_{\mathrm{b}}$, occurs above this regime, further motivated by no clear break seen across the GLEAM bands.} We also require an estimate of the equipartition magnetic field, $B_\mathrm{eq}$. The choice of $B_\mathrm{eq}$ is motivated by Equation 2 of \citet{miley80} and from \citet{jkm+04} we \textcolor{black}{use}, \begin{equation}\label{eq:beq}
B_\mathrm{eq} = 7.91 \left[ \dfrac{1 + k}{\left( 1 +z \right)^{\alpha-3}} \dfrac{S}{\nu^\alpha \theta_x \theta_y l} \dfrac{\nu_\mathrm{max}^{\alpha+\frac{1}{2}} - \nu_\mathrm{min}^{\alpha+\frac{1}{2}}}{\alpha+0.5}\right]^{\frac{2}{7}} \, [\si{\micro\gauss}] \, .
\end{equation}
where $k$ is the relativistic proton--electron energy ratio, $\theta_x$ and $\theta_y$ are the size of the source on the sky in arcseconds, $l$ is the line-of-sight depth, and $\nu_\mathrm{max}$ and $\nu_\mathrm{min}$ are the integration bounds for the luminosity and are chosen to be $\nu_\mathrm{max}=100$~\si{\giga\hertz} and $\nu_\mathrm{min}=0.01$~\si{\giga\hertz}. Here we choose $k$ to be 100 \citep[e.g][]{moffet75} though could be anywhere between 1 and 2000 \citep{pac70}. The choice of $k$ is not overly important as the impact in this range is a change of less than an order of magnitude ($0.9~\si{\micro\gauss} \lesssim B_\mathrm{eq} \lesssim 6~\si{\micro\gauss}$ for $1 \leq k \leq 2000$). We estimate the size of the northern lobe as $\theta_x=570$~arsec, $\theta_y=340$~arcsec on the sky and we assume a line-of-sight depth of $l=127$~\si{\kilo\parsec}. We choose $118$~MHz as the reference frequency, and assume $\alpha=\alpha_\mathrm{inj}$. We estimate $B_\mathrm{eq}\approx2.7~\si{\micro\gauss}$ \textcolor{black}{for the northern lobe, and in the absence of indication of any asymmetry in the environment that would result in lobe asymmetry we make the assumption that the southern lobe has an equivalent magnetic field strength, as we cannot estimate its magnetic field via Eq. \ref{eq:beq} without better knowledge of the extent of the emission (see e.g. Fig. \ref{fig:NGC1534_rgb}, \ref{fig:atca:highres}, and \ref{fig:atca:medres}). We will only report here on the northern lobe fitting results, though for completeness show all fits in Fig. \ref{fig:seds}.} \par

Fig. \ref{fig:seds:north}--\subref{fig:seds:total} show the SEDs of the northern lobe, southern lobe, and total emission, along with model fits. We find that, given $\alpha_\mathrm{inj} = -0.54$ for the northern lobe emission\textcolor{black}{, a total source age is found to be $t_\mathrm{s}=203\pm5$~Myr with an injection time of $t_\mathrm{on} = 44\pm5$~Myr and time since it switched off of $t_\mathrm{off}=158\pm2$~Myr.} Note that quoted errors are simply those from model fitting, and the true uncertainties are much greater as many assumptions are made in this process. \textcolor{black}{Notably, our value of $t_\mathrm{s}$ suggests a break frequency of $\sim 502$~MHz \citep[see e.g. Equation 1 of][]{al87}, above the GLEAM frequency coverage validating our choice of $\alpha_\mathrm{inj}$ found from those data.} \textcolor{black}{As discussed in \citet{har17}}, these times should be considered as \textcolor{black}{`characteristic'.} From the $\mathrm{CI}_\mathrm{off}$ model of the northern lobe, we estimate the 1.4-GHz flux density as $S_{1.4}^{\mathrm{north}} \approx 41$~\si{\milli\jansky}. Assuming the source size is the same as at the 1.51-GHz size of $\sim60$~arcmin$^2$, then the surface brightness is $\sim0.7$~mJy$\,$arcmin$^{-2}$. Assuming NGC~1534 is the original host, and assuming the true emission is represented by a symmetric set of lobes of flux density $2\times S_{1.4}^{\mathrm{north}}$, the core-to-lobe luminosity ratio is $P_{1.4}^{\mathrm{core}}/P_{1.4}^\mathrm{lobe} \approx 0.02$. \par

Assuming the host of the emission is NGC~1534 and with a distance of 230~kpc from northern lobe centre (i.e. the equivalent hotspot) to NGC~1534, the minimum velocity of the lobe must be $\sim0.014c$, which is on the same order of magnitude \textcolor{black}{as FR-II \citep{fr74} sources} \citep[e.g.][]{lpr92}.\par

One should be cautious comparing integrated flux densities of maps with different $u$--$v$ coverage. As there are differences between not only the MWA and ATCA observations, but also the Molonglo and ATCA observations, we may be biasing the spectrum to be much steeper above 1~GHz. This high-frequency, steep-spectrum bias suggests the $t_{\mathrm{off}}$ estimate is an upper limit, as the true age will be younger with a flatter high-frequency spectrum. Further ATCA observations to fill in the $u$-$v$ plane would be required to confirm this. We note that \citet{hje+15} find a significantly higher integrated flux density at 185-MHz than what is suggested here, however, subsequent improvements to the MWA primary beam model and general flux scale used by \citet{gleamegc} can account for this discrepancy. 
\subsection{Polarimetry}\label{sec:polarimetry}

The ATCA operates with linear XX, YY, XY, and YX cross-correlations which allow measurement of Stokes $I$, $Q$, $U$, and $V$, thus we investigate the polarisation properties of the emission surrounding NGC~1534. We are interested in the linear polarization defined via Stokes $Q$ and $U$, \begin{equation}
P = Q + iU \, ,
\end{equation}
with, assuming no circularly polarized emission (Stokes $V$), the total polarization intensity \begin{equation}
\|P\| = \sqrt{Q^2 + U^2} \, .
\end{equation}

\subsubsection{Rotation measure synthesis}\label{sec:rm}
As the 2.1-GHz data have a large fractional bandwidth and reasonably small channels, we perform rotation measure synthesis \citep[RM-synthesis;][]{bd05}---a method to investigate rotation measure on a non-contiguous spectrum, building on the rotation measure work of \citet{bur66}. As in \citet{bur66}, \citet{bd05} define the Faraday depth, $\phi$, via \begin{equation}\label{eq:faraday_depth}
\phi(\boldsymbol{r}) = 0.81 \displaystyle\int\limits_{\mathrm{there}}^{\mathrm{here}} n_{\mathrm{e}} \boldsymbol{B} \cdot \mathrm{d}\boldsymbol{r} \quad [\si{\radian\per\m\squared}]\,,
\end{equation}
where $n_\mathrm{e}$ is the electron density in \si{\per\centi\m\squared}, $\boldsymbol{B}$ is the intervening magnetic field in \si{\micro\gauss}, and $\mathrm{d}\boldsymbol{r}$ is an infinitesimal path length in \si{\parsec}. Though we cannot usually measure the electron density, the sign of the Faraday depth gives the average magnetic field direction, where a negative $\phi$ is given by a magnetic field in the direction of the observer. Additionally, the Faraday depth spectrum may show other sources along the line of sight. Intrinsic source rotation measure is defined as \begin{equation}\label{eq:rm}
R\!M_{0} = \dfrac{\mathrm{d}\chi_{\mathrm{P}}}{\mathrm{d}{\lambda_{\mathrm{obs}}}^2} \left(1 + z \right)^2  - R\!M_{\mathrm{gal}} - R\!M_{\mathrm{other}}\quad [\si{\radian\per\m\squared}]\,,
\end{equation}
where $\lambda_{\mathrm{obs}}$ is the observed wavelength, and $z$ is the redshift of the source in question, $R\!M_{\mathrm{gal}}$ is RM contribution from Galactic Faraday rotation and $R\!M_{\mathrm{other}}$ is the RM contribution from other foreground or background sources. The rotation measure gives insight into the source magnetic field as well as any intervening or background magnetic field sources. \par
For RM-synthesis, we use the EW367 ATCA data and create Stokes $Q$ and $U$ cubes with axes $\alpha_{\mathrm{J2000}}$, $\delta_{\mathrm{J2000}}$, and $\nu$, where $\nu$ represents a single channel of 1~\si{\mega\hertz}. Only the EW367 observation is used as it had the least flagging due to RFI which enabled a larger fractional bandwidth/more individual channels to be used at only a small loss to sensitivity and $u$--$v$ coverage. Each plane in the cube, corresponding to a 1~\si{\mega\hertz} channel of the original 2.1-\si{\giga\hertz} data, are imaged to the same dimensions---no CLEANing is done on the 1-MHz images. Primary beam corrections are applied for each plane at the given frequency, though pixels outside the FWHM of the primary beam of the highest frequency 1-MHz image are blanked in the output FITS cubes, thus we do not expect noise to vary significantly across the planes. The imaging is done on a per-pointing basis, with each plane a mosaic using \textsc{linmos} as in Section \ref{sec:highres}. For channels where all data are flagged, we skip those in the cube-forming/imaging process and move on to the next channel.

\begin{figure}
\begin{subfigure}{1\linewidth}
\includegraphics[width=1\linewidth]{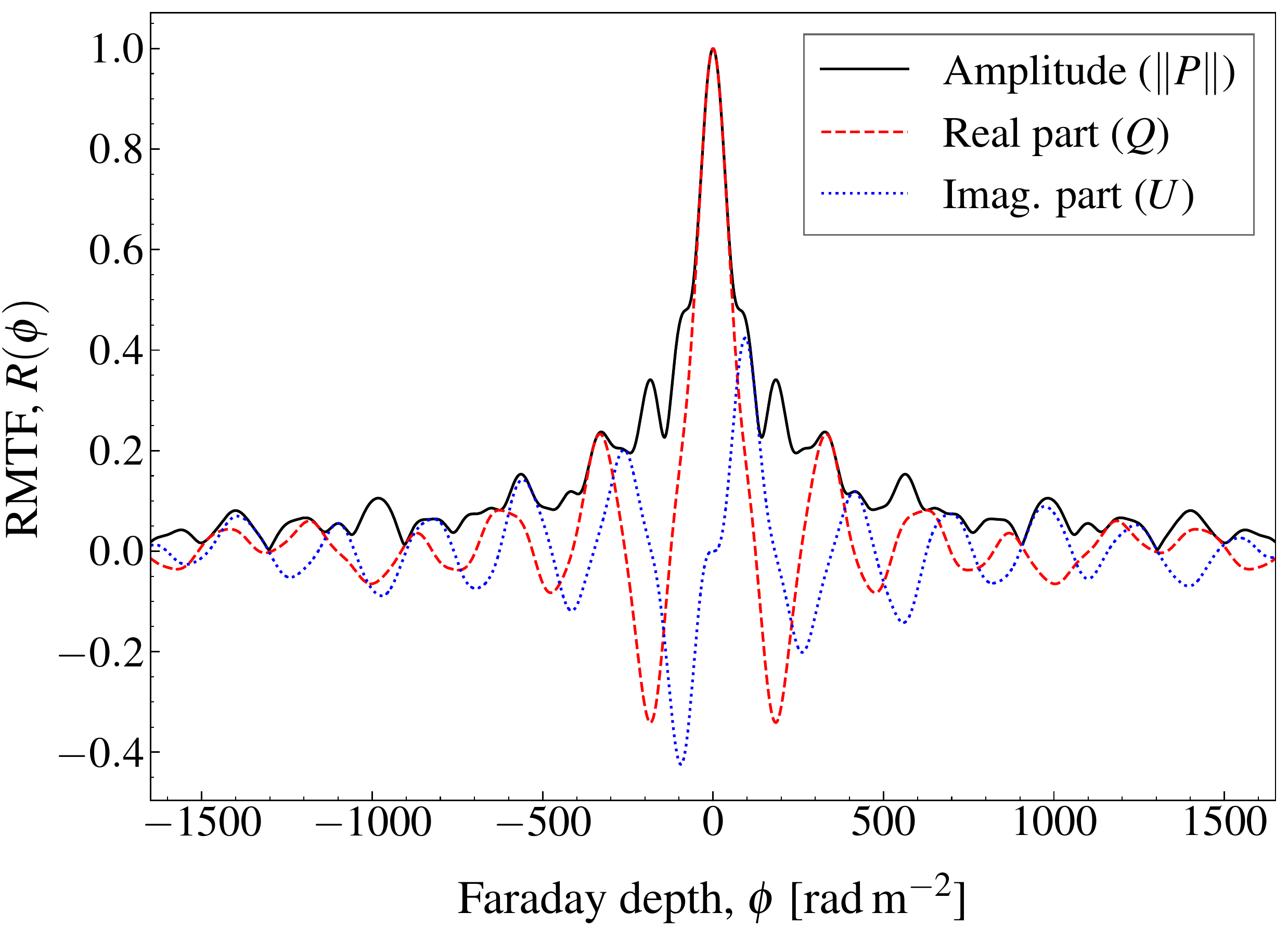}
\caption{\label{fig:rm:rmtf}}
\end{subfigure}\\%
\begin{subfigure}{1\linewidth}
\includegraphics[width=1\linewidth]{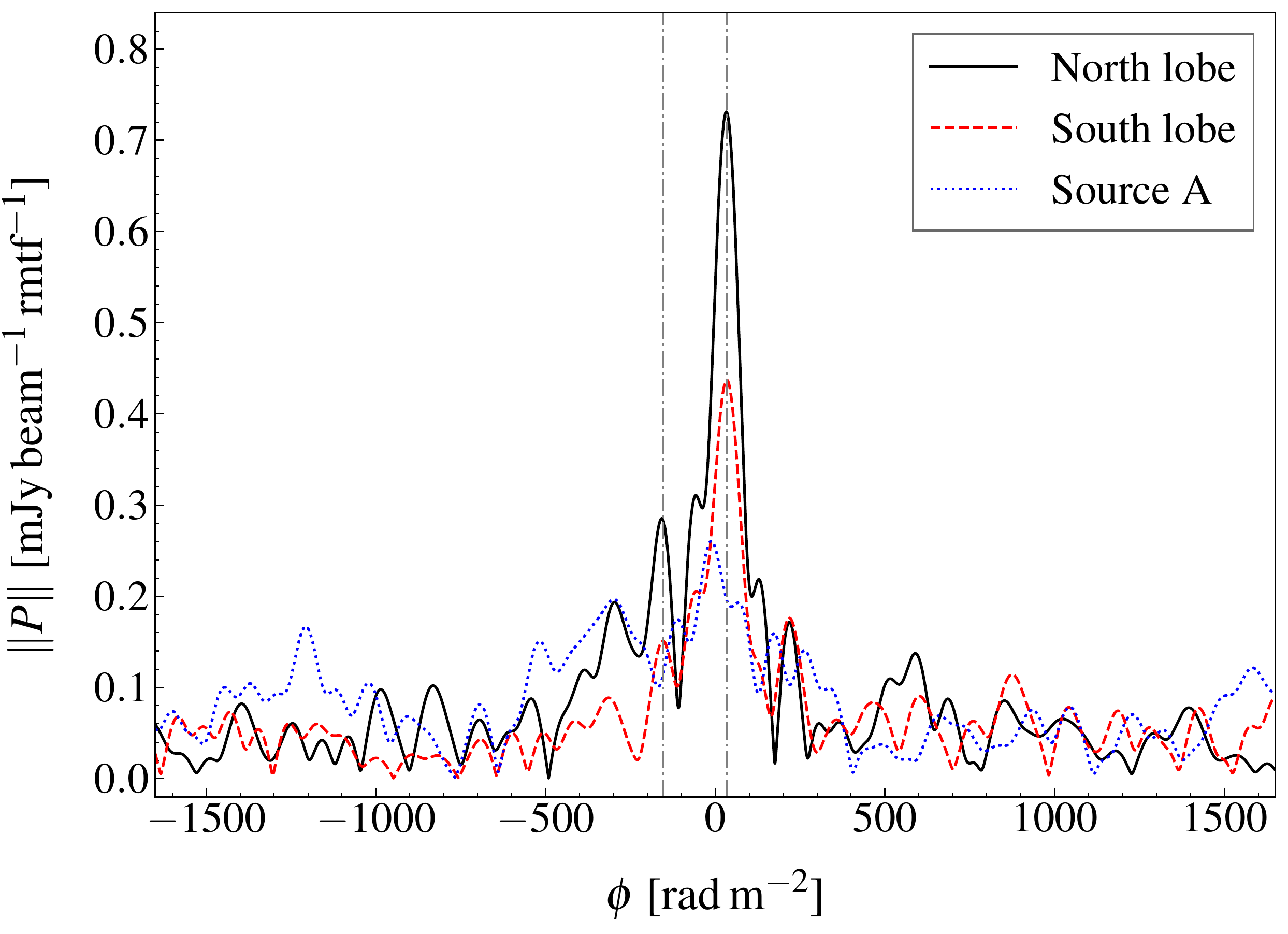}
\caption{\label{fig:rm:NSA}}
\end{subfigure}%
\caption{\subref{fig:rm:rmtf}: the RMTF between $-1650 \leq \phi \leq 1650$. \subref{fig:rm:NSA}: the polarised intensity along the Faraday depth cube of 3 pixels corresponding to a pixel within the northern lobe (black, solid---$04\hour08\min29\fs7, -62\degr38\arcmin56\farcs8$), within the southern lobe (red, dashed---$04\hour09\min39\fs7, -62\degr58\arcmin08\farcs9$), and Source A (blue, dotted---$04\hour08\min42\fs0, -62\degr33\arcmin00\farcs8$). Marked with vertical lines are peaks of interest in the Faraday depth spectrum. In both panels, the resolution in $\phi$ is 1~\si{\radian\per\m\squared}. The vertical lines represent the two detected RM features at $-153$ and $+33$~\si{\radian\per\m\squared}. }
\label{fig:rmtf}
\end{figure}

The $Q$ and $U$ cubes, along with a list containing frequencies for each plane,  are then used by the RM-synthesis code developed by M.~A. Brentjens \footnote{\url{https://github.com/brentjens/rm-synthesis}} to generate a Rotation Measure Transfer Function (RMTF, also known as the rotation measure synthesis function, shown in Fig. \ref{fig:rm:rmtf}) and cube of $\alpha_{\mathrm{J2000}}$, $\delta_{\mathrm{J2000}}$, and Faraday depth $\phi$ in units of the Faraday dispersion function, $F(\phi)$. The resolution chosen for synthesising RM is 1~\si{\radian\per\m\squared}. We synthesised the Faraday dispersion in the range $-1650 \leq \phi \leq +1650$,  The top panel of Fig. \ref{fig:rmtf} shows the RMTF. The polarised intensity, $\|P\|$ in \si{\jansky\per\beam\per\rmtf}, is equal to the Faraday dispersion function, $F(\phi)$, in the case of sources that are discrete in $\phi$ \citep{db05}. \par

\begin{figure*}
\centering
\begin{subfigure}{0.5\linewidth}
\includegraphics[width=1\linewidth]{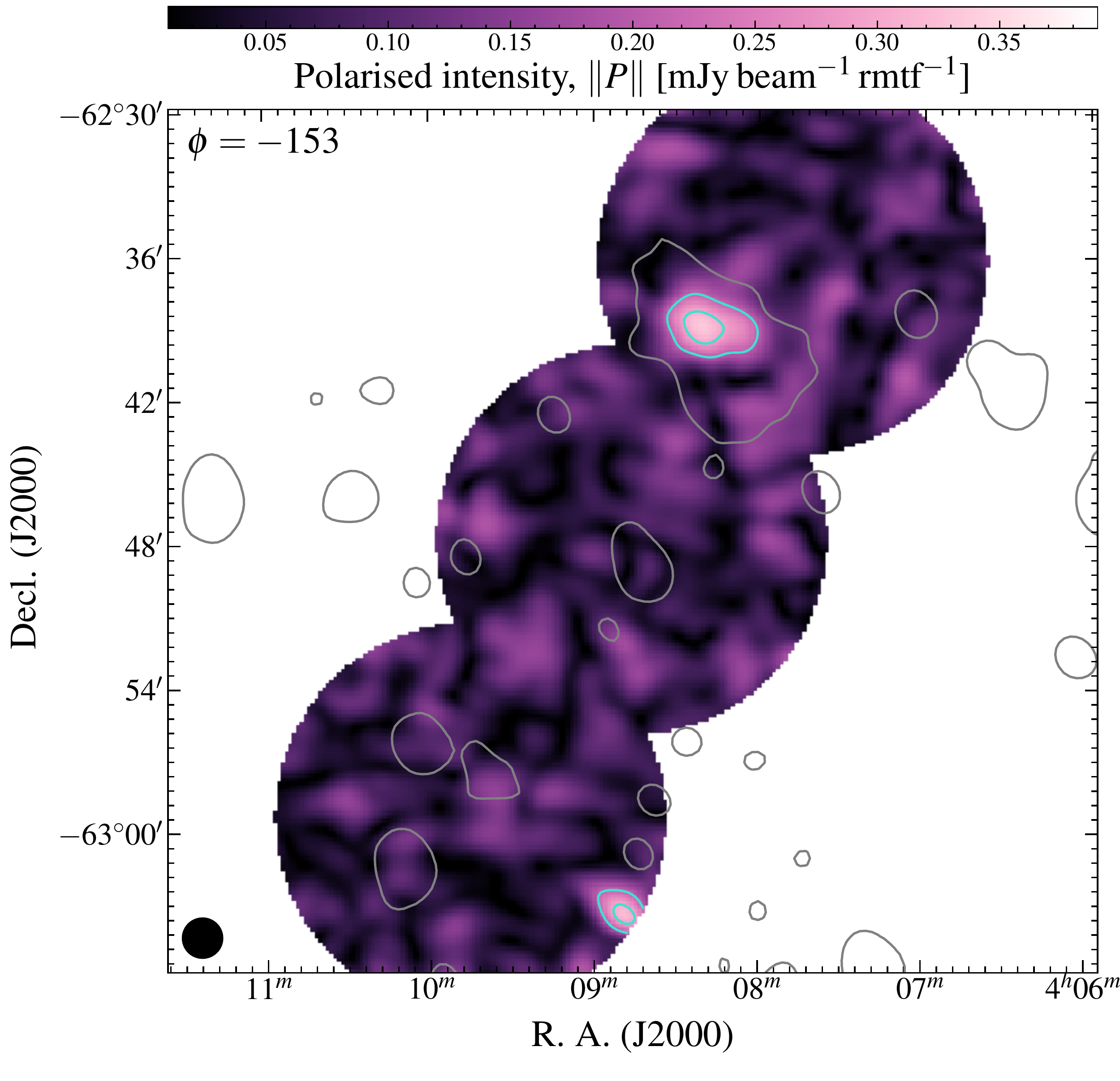}
\caption{\label{fig:rm:1497}}
\end{subfigure}%
\begin{subfigure}{0.5\linewidth}
\includegraphics[width=1\linewidth]{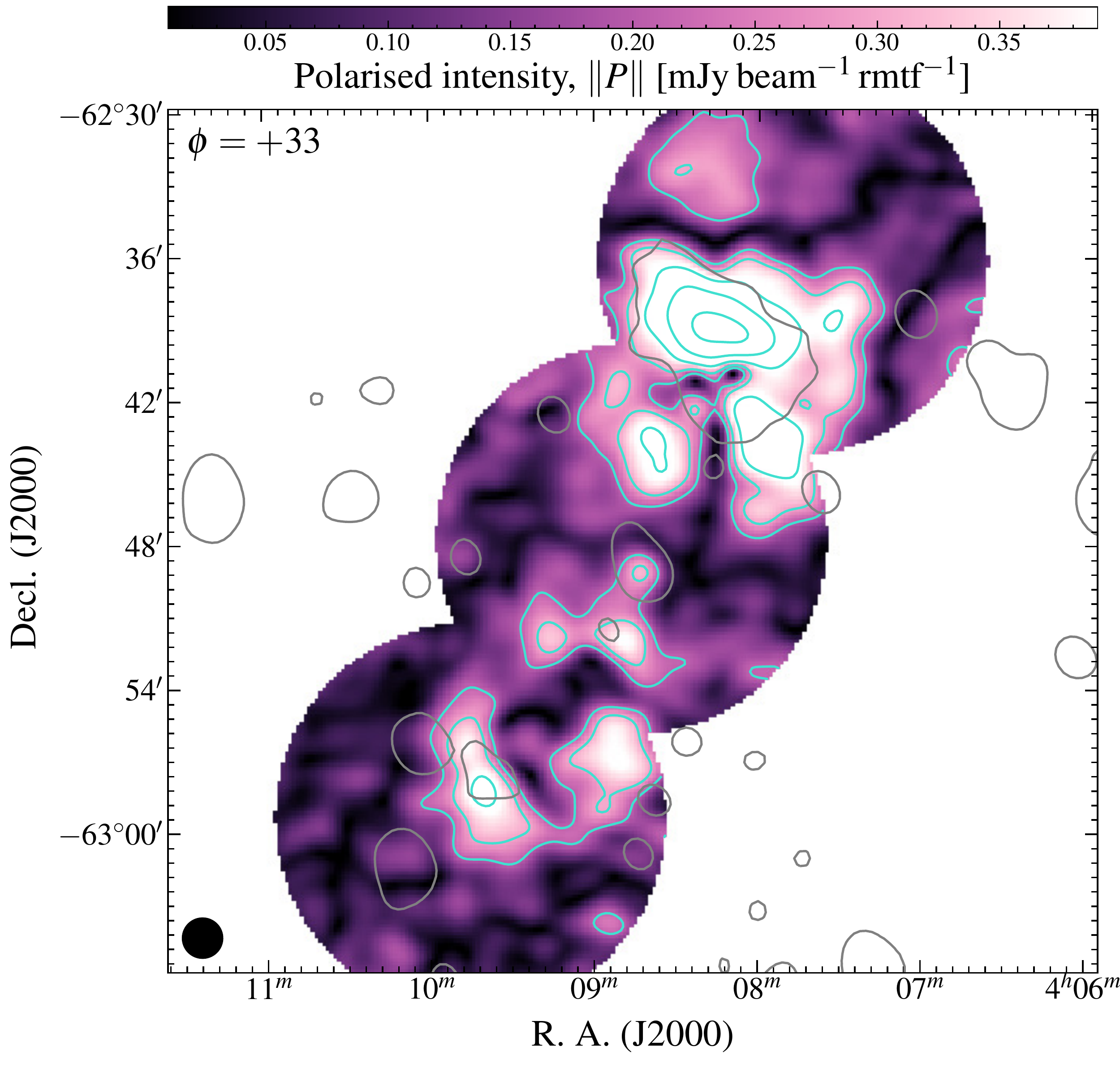}
\caption{\label{fig:rm:1684}}
\end{subfigure}%
\caption{Planes in the Faraday depth cube as indicated in Fig. \ref{fig:rm:NSA}. \subref{fig:rm:1497}: $\phi=-153$. \subref{fig:rm:1684}: $\phi=+33$. In both panels, the single black contour is of the medium-resolution 1510-MHz image at 810~\si{\micro\jansky\per\beam}. The turquoise contours are the linear polarization intensity at the specific Faraday depth, beginning at 210~\si{\micro\jansky\per\beam\per\rmtf} and increasing with factors of $\sqrt{2}$. The black ellipse in the lower-left corner is the beam shape of the Faraday depth cube. Both images share the same linear colour scale.}
\end{figure*}

Fig. \ref{fig:rm:NSA} shows the Faraday depth spectra of three representative pixels: within the northern lobe ($04\hour08\min29\fs7, -62\degr38\arcmin56\farcs8$), the southern lobe ($04\hour09\min39\fs7, -62\degr58\arcmin08\farcs9$), and Source A ($04\hour08\min42\fs0, -62\degr33\arcmin00\farcs8$). The north and south lobe pixels have peaks at a Faraday depth of $+33$ and $+34$~\si{\radian\per\m\squared}, respectively, and Source A shows a peak at $-12$~\si{\radian\per\m\squared}. The main peak in the north and south lobes is close to the estimated Galactic foreground RM of $+27$~\si{\radian\per\m\squared} \citep[][but see also \citealt{ojr+12}]{ojg+15}. This Galactic foreground value is taken from an average value within 1000 arcsec of NGC~1534, which comprises approximately four pixels of the \textsc{HEALPix} image of the Galactic Faraday depth produced by \citet{ojg+15}. Fig. \ref{fig:rm:1684} shows the plane in the Faraday depth cube at $\phi=+33$, showing large-scale emission beyond the size of the emission from NGC~1534, further suggesting Galactic (or otherwise foreground) origin rather than the intrinsic magneto-ionic plasma of NGC~1534's lobes. Fig. \ref{fig:rm:1497} shows the second isolated peak in the Faraday depth spectrum of the northern lobe pixel (marked in Fig. \ref{fig:rm:NSA}) at $-153$~\si{\radian\per\m\squared}. We do not have enough information about the intergalactic medium to know with 100 per cent certainty whether this peak corresponds to a non-Galactic screen external to the radio plasma, or to the radio lobe itself. However, as the position corresponds to the peak brightness of the polarised emission in the lobes, it is likely to be associated with the radio galaxy itself.

\subsubsection{Continuum polarimetry}\label{sec:atca_pol}

\begin{figure*}
\begin{subfigure}{0.5\linewidth}
\includegraphics[width=1\linewidth]{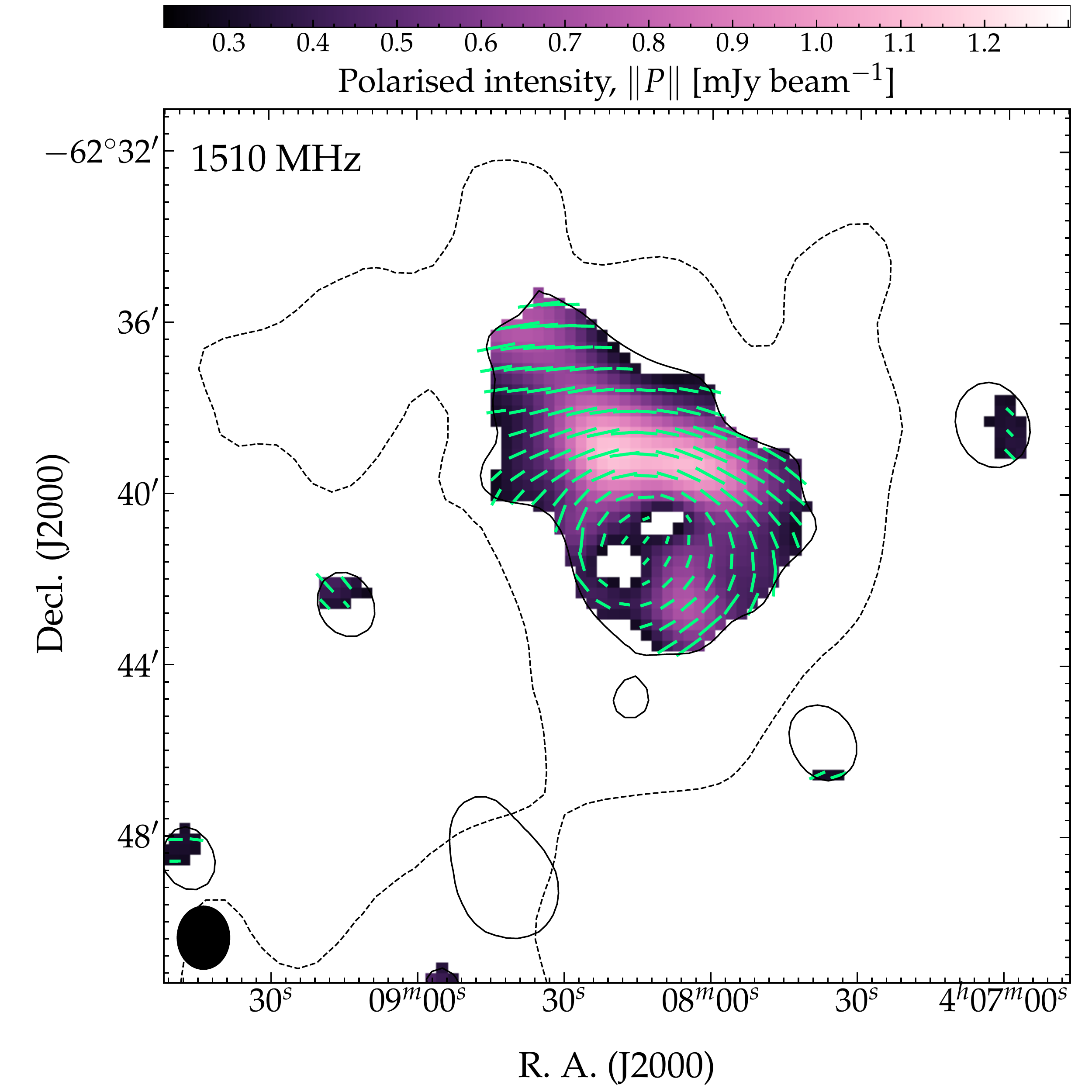}
\caption{\label{fig:atca_pol:1510}}
\end{subfigure}\hfill%
\begin{subfigure}{0.5\linewidth}
\includegraphics[width=1\linewidth]{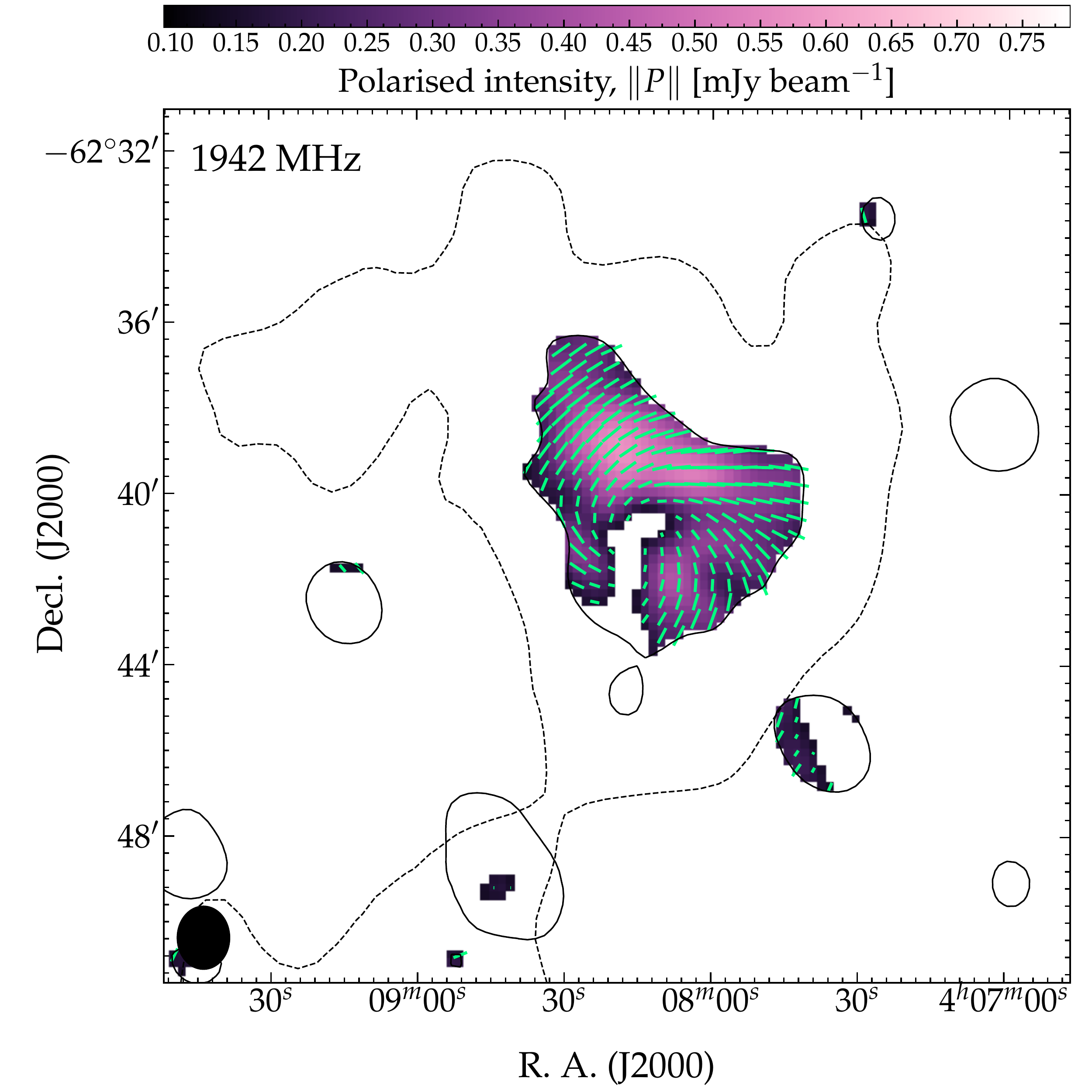}
\caption{\label{fig:atca_pol:1942}}
\end{subfigure}%
\caption{Polarization images. \subref{fig:atca_pol:1510}: 1510-MHz subband image. \subref{fig:atca_pol:1942}: 1942-MHz subband image. The background in both panels is the total linear polarization intensity map (i.e. $\|P\|=\sqrt{Q^2 + U^2}$) which is overlaid with a single black, dashed GLEAM 200-MHz contour at $3\sigma_{\mathrm{rms}}$ and a single black, solid ATCA Stokes $I$ contour at $3\sigma_{\mathrm{rms}}$ of the medium-resolution image. The fields are the $B$-field and the vector lengths are proportional to $m_P$ where 5 pixels correspond to $m_{P} = 1$. The position angles are corrected for Galactic Faraday rotation, assuming $\phi_{\mathrm{gal}}=+33$~\si{\radian\per\m\squared}, and an additional Faraday screen at $\phi=-153$~\si{\radian\per\m\squared}.}
\label{fig:atca_pol}
\end{figure*}

We follow a similar imaging procedure as in Section \ref{sec:lowres} (i.e., without antenna 6). The Stokes $I$ images are deconvolved in a similar manner to the Stokes $I$ images of Section \ref{sec:lowres}, though for the Stokes $Q$ and $U$ images we use the complex implementation of the Steer-Dewdney-Ito \citep[SDI;][]{sdi84} CLEAN algorithm offered by the task \textsc{cclean}  \citep[][]{pj16}. The SDI CLEAN algorithm is better at CLEANing extended sources than the traditional Hogb\"{o}m \citep{hog74} or Clark \citep{cla80} CLEAN algorithms as used by \textsc{mfclean}. Complex CLEAN acts on both Stokes $Q$ and $U$ in a dependent fashion. As linear polarisation, $P$, is a complex quantity, the complex CLEAN algorithm properly accounts for this complex vector nature of the signal. We produce total polarisation intensity maps ($\|P\|$), shown in Fig. \ref{fig:atca_pol}, overlaid with vectors of magnitude proportional to the fractional polarisation, $m_P = \|P\|/I$, and directions representing the apparent magnetic field, $\chi_P + \pi/2 - R\!M\lambda^2$, where $\chi_P$ is the electric vector position angle defined via 
\begin{equation}
\chi_P = \frac{1}{2}\arctan \frac{U}{Q} \, ,
\end{equation} 
and $R\!M$ is the total line-of-sight RM. In making polarization images (intensity, fractional polarization, and position angle) we use a $3\sigma_{\mathrm{rms},QU}$, $3\sigma_{\mathrm{rms},I}$ cut to the intensity and a $3\sigma_\mathrm{rms} = 10$~degree cut to the position angle. This results in no detected polarised emission from the 2375- and 2807-MHz bands. Fig. \ref{fig:atca_pol:1510} and \ref{fig:atca_pol:1942} show the polarisation intensity maps for the 1510- and 1942-MHz bands, respectively, with vectors of magnitude defined by the factional polarisation and magnetic field directions. We de-rotate the position angles based on an assumed Galactic Faraday depth of $+33$~\si{\radian\per\m\squared} and for the additional peak at $-153$~\si{\radian\per\m\squared} (see Section \ref{sec:rm}). The average fractional polarisation across the source in the 1510- and 1942-MHz bands is $42\pm14$ and $43\pm14$~\%, respectively, and in the higher bands little polarization is detected, following the Stokes $I$ images. The field directions appear curled which is not typically seen in the intrinsic magnetic fields of radio galaxy lobes \citep[e.g.][]{bp84}, unless the lobe is bent or twisted \citep[e.g.][]{lbp+08}, however, the curling seen here would require the northern lobe to have fallen completely back in on itself. \par

\section{Discussion}
NGC~1534 represents the rare chance to study a relatively near-by radio galaxy with diffuse low-surface brightness emission. It is peculiar for a number of reasons including its position in the field, not a cluster, and the seeming discrepancy between the position of the optical galaxy and typical position for radio emissions appears to be slightly offset. Here we consider the properties of NGC~1534 and the environment that surrounds it and argue that it is not inconsistent that if the AGN has switched off, we could see a drift in the position of the source in the loose group environment which could account for the slight misalignment of optical host and radio jet positions.  

\subsection{Radio emission from disk galaxies in the NGC~1534 field}

\begin{figure*}
\centering
\begin{subfigure}{0.355\linewidth}
\includegraphics[width=1.0\linewidth]{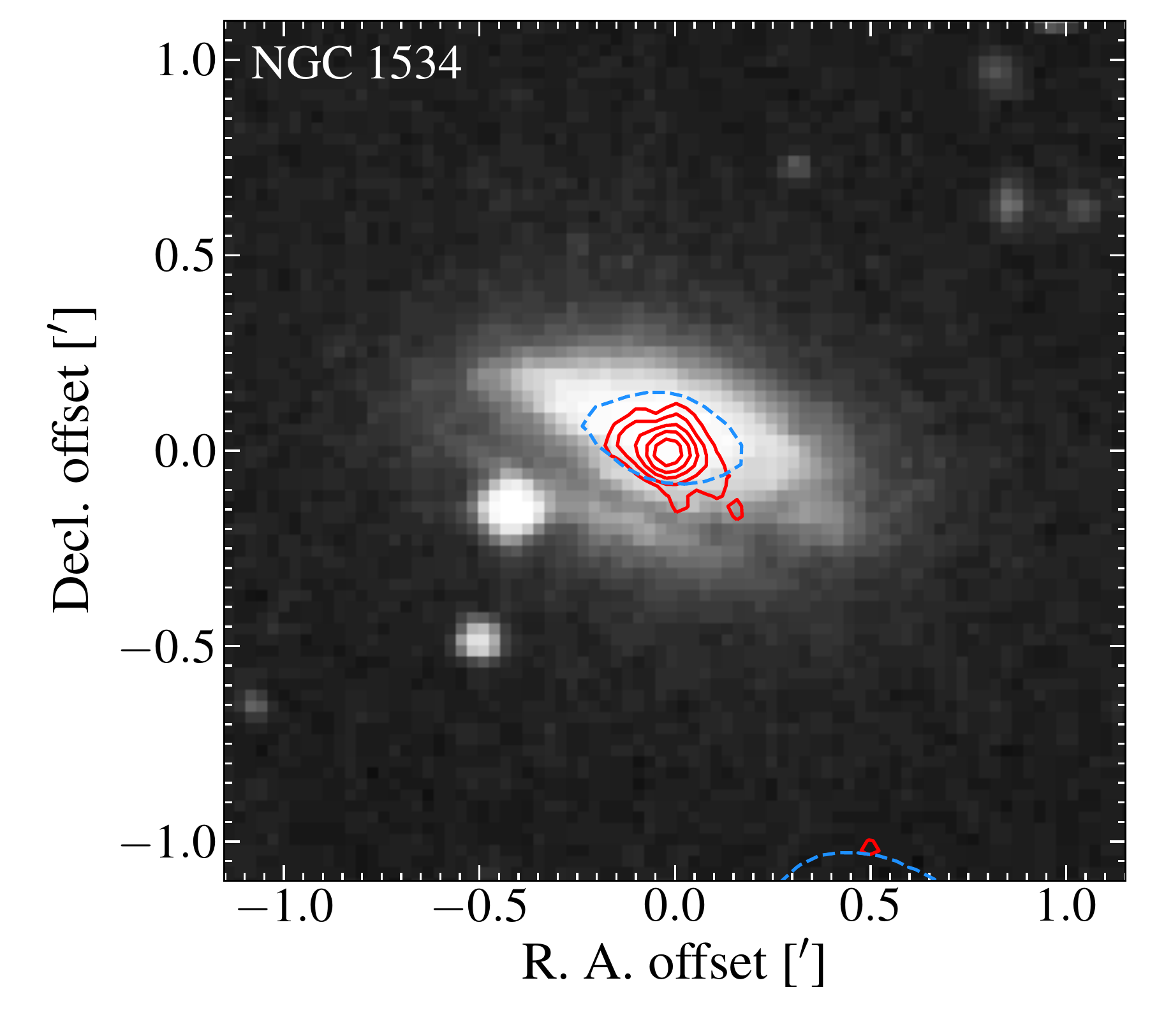}
\caption{\label{fig:disk_galaxies:NGC1534}}
\end{subfigure}\hfill%
\begin{subfigure}{0.315\linewidth}
\includegraphics[width=1.0\linewidth]{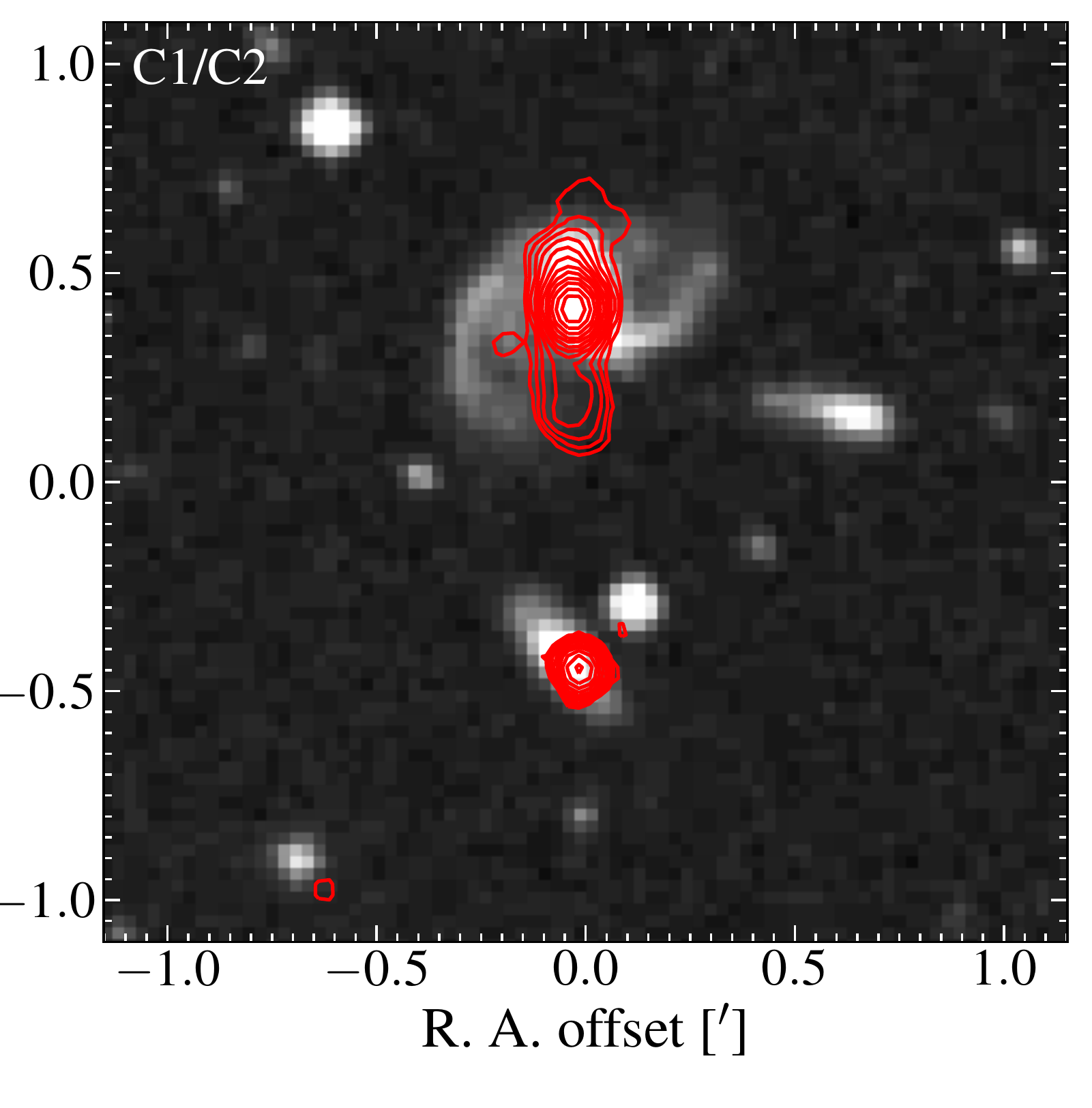}
\caption{\label{fig:disk_galaxies:C1C2}}
\end{subfigure}\hfill%
\begin{subfigure}{0.315\linewidth}
\includegraphics[width=1.0\linewidth]{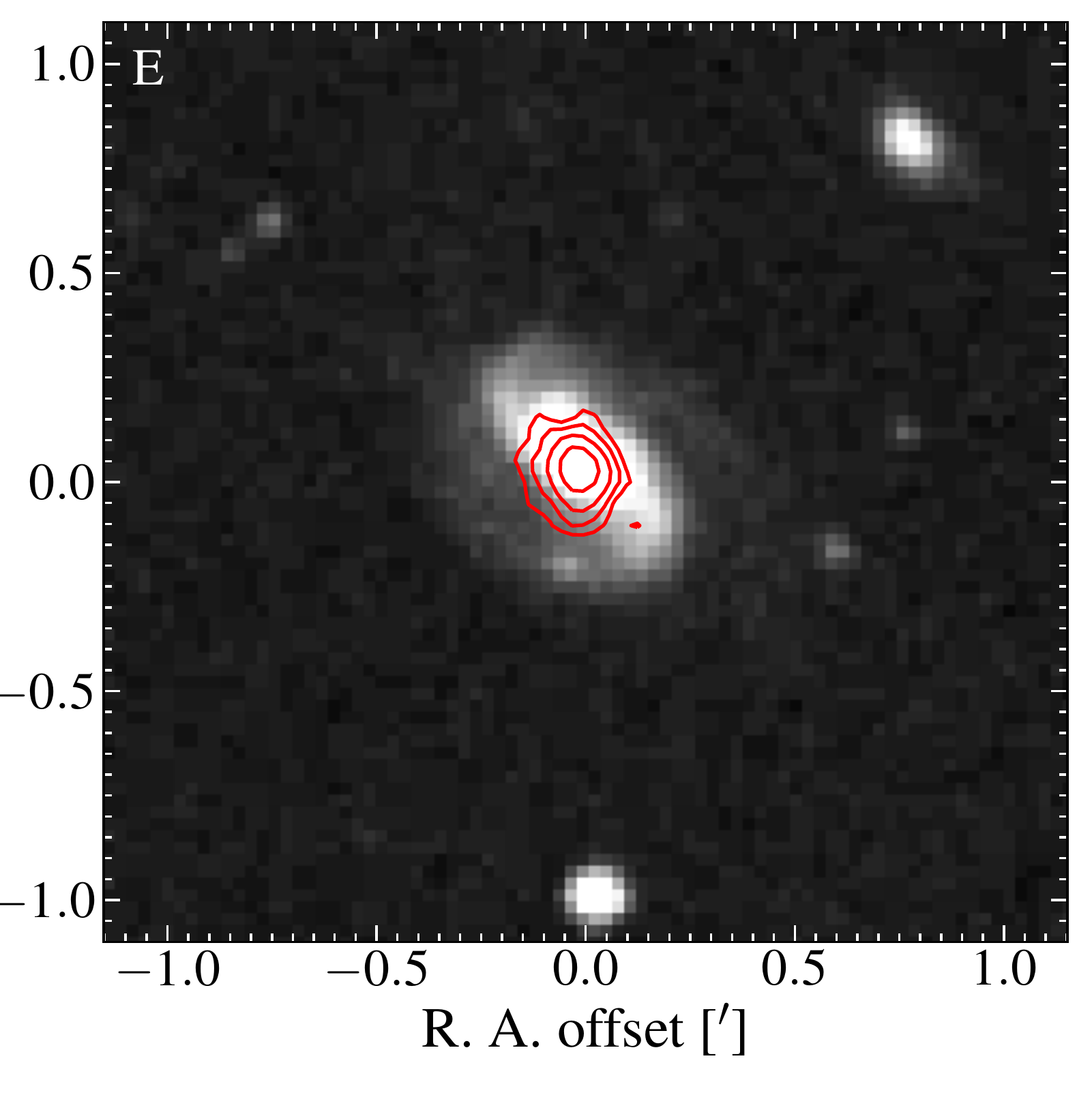}
\caption{\label{fig:disk_galaxies:E}}
\end{subfigure}%
\caption{Disk galaxies in the NGC~1534 field with significant radio emission at 2.2~GHz. \subref{fig:disk_galaxies:NGC1534}: NGC~1534. \subref{fig:disk_galaxies:C1C2}: Sources C1 and C2. \subref{fig:disk_galaxies:E}: Source E. The background images are the blue band UKST images, and the solid, red contours are from the 2.2-GHz wideband ATCA image, beginning at $3\sigma_\mathrm{rms}$ for NGC~1534 and $4\sigma_\mathrm{rms}$ for Sources C1/C2 and E. The dashed, blue contour in \subref{fig:disk_galaxies:NGC1534} is the 17-GHz medium-resolution image at $3\sigma_\mathrm{rms}$.}
\label{fig:disk_galaxies}
\end{figure*}

Radio emission is detected in four disk galaxies in the NGC~1534 field, including NGC~1534 itself. Fig. \ref{fig:disk_galaxies:NGC1534}--\subref{fig:disk_galaxies:E} show the relevant galaxies with 2.2-GHz contours overlaid. NGC~1534 and Source E show typical diffuse emission consistent with low nuclear activity, however both Source C1 and C2 have reasonably strong nuclear activity, with Source C1 showing extensions north and south. Sources C1 and C2 are reported as part of the galaxy triple AM~0409$-$630 \citep[$z=0.0481\pm0.0002$;][]{shd+92}, though no redshift is directly available for Source C2. At this redshift, the extension seen in the radio structure of Source C1 are of the order 10~kpc. Additionally, further extension in the radio emission traces the central portion of the bar. Source E is part of the triple AM~0406$-$624, though itself has no redshift. Its triple members, PGC~014488 and LEDA~075047 have redshifts $z=0.0189\pm0.0002$ and $z=0.0193\pm0.0002$ \citep{jrs+09}, respectively. If the triple association is correct, then we estimate a redshift of $z\sim0.0191$. At this redshift, the separation from NGC~1534 is only $\sim380$~\si{\kilo\m\per\s}. This implies a loose group association. 

Importantly, we see from Figures \ref{fig:atca_examples} and \ref{fig:disk_galaxies} that the radio emission from the core NGC~1534 is extended and diffuse, even at high frequencies. It is therefore very likely to be the result of only star formation in the disk of the galaxy, and there is no significant AGN present in the NCG~1534 core.

\subsection{Star formation in NGC~1534}

As mentioned above, our ATCA images at 2.1 and 17 \si{\giga\hertz} clearly show that the main component of radio emission from NGC~1534 is extended and diffuse, likely originating within the disk from star formation rather than from any significant nuclear activity. We can test the validity of this assertion by comparing the mid-infrared star-formation rate (SFR) estimators via \textcolor{black}{\emph{Widefield Infrared Survey Explorer} \citep[\emph{WISE};][]{wise,neowise}} measurements \citep[e.g.][]{jmt+13} with the 1.4-\si{\giga\hertz} SFR estimator via our 2.1-\si{\giga\hertz} ATCA measurements \citep[][]{con92,hmcr98,chmr98}. At present, one must be cautious when using radio luminosity a SFR estimator as doing so assumes all radio emission is directly related to past star formation and not AGN or other nuclear activity. In the case of a normal radio galaxy with typical double-lobed structure, the AGN component will dramatically dominate the comparatively feeble radio emission from supernova remnants thus providing a massively overestimated SFR. Likewise, infrared SFR estimators will run into similar problems, though much less dramatically. \par
\begin{table}
\caption{Mid-infrared properties of NGC~1534 from \emph{WISE} \citep{allwise}. For these data we estimate a spectral index of 0.2 for obtaining correction factors for magnitude to flux density conversion and for estimating luminosity, $L_\nu$. The Solar luminosity, $\mathrm{L}_{\odot}$, is taken to be $3.839 \times 10^{33}${}~\si{\watt} as per \citet{jmt+13}.}
\label{tab:wise}
\centering
\resizebox{\linewidth}{!}{\begin{tabular}{l c c c c}
\hline\hline
Band & $\lambda$ & $m_{\mathrm{Vega}}$ & $S_\nu$ & $\nu L_\nu$ \\
     & (\si{\micro\m}) & (\si{\mag}) & (\si{\milli\jansky}) & ($\times 10^9 \, \mathrm{L}_{\odot}$) \\
\hline
$W1$ & 3.4             & $10.365 \pm 0.022$ & $22.12 \pm 0.45$ & - \\  
$W2$ & 4.6             & $10.377 \pm 0.019$ & $12.14 \pm 0.21$ & - \\
$W3$ & 12              & $8.193  \pm 0.019$ & $16.73 \pm 0.29$ & $0.763 \pm 0.013$ \\
$W4$ & 22              & $6.567  \pm 0.050$ & $19.75 \pm 0.91$ & $0.491 \pm 0.023$ \\
\hline
\end{tabular}}
\end{table}
\textcolor{black}{Fortunately, NGC~1534 is detected in all four bands of the \emph{WISE} all-sky catalogue, AllWISE \citep{allwise}. \emph{WISE} has bands $W$1--4, corresponding to wavelengths $3.4$, $4.6$, $12$, and $22$ \si{\micro\m}. Table \ref{tab:wise} presents $W$1--4 band Vega-calibrated magnitudes as well as flux densities and luminosities assuming a spectral index across the $W$1--4 bands of 0.2.} With comparison to Figure 11 of \citet{ydt+13}, we can see from \emph{WISE} colours $W1 - W2 = -0.012$ and $W2 - W3 = 2.184$ that NGC~1534 may indeed have some form of nuclear activity, though may also be considered as star-forming. With reference to Equation 2 of \citet{jmt+13}, we estimate the SFR from the $W4$ measurement: $S\!F\!R_{22} = 0.37 \pm 0.04$ M$_{\odot}$\,\si{\per\year}. Note that the \emph{WISE} $W4$ band traces mostly the warm interstellar medium dust emission and is a lower estimate of the full SFR. \par
For the 1.4-\si{\giga\hertz} SFR, we use the 1.4-GHz power of NGC~1534 derived in Section \ref{sec:interlopers}. We use the 1.4-\si{\giga\hertz} SFR estimator of \citeauthor{hop98} (\citeyear{hop98}, Equation 7.1; adapted from \citealt{con92}) to estimate $S\!F\!R_{1.4} = 0.38 \pm 0.03$ M$_{\odot}$\,\si{\per\year}, which is valid for the non-thermal synchrotron component, mostly tracing supernova remnants and is valid for stars with $M\geq5$~M$_\odot$. This value is in good agreement to $S\!F\!R_{22}$ derived from \emph{WISE} results. The agreement in these two estimators suggests a lack significant emission from an AGN and further suggests NGC~1534, if the original host of the radio lobes, is no longer fuelling them.

\subsection{A dead radio galaxy}\label{sec:host}

\citet{hje+15} consider NGC~1534 to be a previously active radio galaxy, with the emission seen surrounding it the remnants of the ancient lobes. The ATCA observations of NGC~1534 point toward a lack of AGN activity, which, coupled with the offset of NGC~1534 from the centre of the lobes, suggests a dying---or dead---radio galaxy. Such an object is less common outside of galaxy clusters \citep[e.g.][]{cor87,mpm+11}. The spectral properties of the lobes of NGC~1534 are consistent with the cluster-based dead radio galaxy sample of \citet{mpm+11} as well as non-cluster based examples \citep[e.g.][]{jkm+04,pmd+07,bgm+16}. However, in dense cluster environments the radio plasma becomes confined and the lobe size also remains much smaller \citep{mpm+11}. In the case of NGC~1534, we have emission $\gtrsim 600$~kpc outside of any dense environment which provides an interesting example of a dead field radio galaxy. 

\textcolor{black}{The emission surrounding NGC~1534 has many similarities with the remnant emission `blob1' \citep{bgm+16}---namely its location in an underdense environment, its projected size, and the spectral energy distribution. The $t_\mathrm{off}/t_\mathrm{s}$ ratio of its `characteristic time', 0.8, is also similar, however, the spectral age determined through CI$_\mathrm{off}$ modelling may be poorly represented \citep[see e.g.][]{har17}. Assumptions such as a constant magnetic field over the lifetime of the source or a constant spectrum over the extent of the source can lead to uncertain spectral ages \citep[see e.g.][]{har17,hhm+17,trsk18,tsk18}, and adiabatic losses will result in an underestimate to the dynamical age \citep[particularly in an underdense medium;][]{br00}. Work has been done to incorporate adiabatic losses into radio galaxy spectral models \citep[see][]{gmb17,hcm+18}, and model fitting across the extent of a resolved source can alleviate the issues with integrated spectra \citep[e.g.][and \citealt{brats1,brats2} with the use of \textsc{brats} in this regard]{har17}. Additionally, \citet{tsk18} discuss how the CI spectrum can be modelled independent of the magnetic field, though note that the magnetic field strength is required for an estimate of the source's synchrotron age.} 
%Despite this, the ratio $t_\mathrm{off}/t_\mathrm{s}\approx0.78$ is consistent with the range found by \citet{mpm+11} and the total time is not the most extreme spectral age found of such sources \citep[e.g. WNB1829$+$6911 has $t_\mathrm{s}=218$~Myr, though has large uncertainties;][]{mpm+11}.
 Interestingly, we see that the intrinsic magnetic field of dead radio galaxies does not differ much based on environment \textcolor{black}{(assuming a fixed magnetic field strength)}---from a mixture of cluster and non-cluster sources, for both the samples of \citet{pmd+07} and \citet{mpm+11}, a mean equipartition magnetic field of $\sim13$~\si{\micro\gauss} exists, though note that their equipartition calculations use a fixed energy range rather than a fixed frequency range as used here, and are up to a factor of two larger. Additionally, `blob1' has 1~\si{\micro\gauss} \citep{bgm+16}, emission surrounding NGC~5580 and NGC~5588 in a poor group has 2.5~\si{\micro\gauss} \citep[][though note the authors are uncertain of its classification]{diw+14}, and the archetypcal B2~0924$+$30 in the poor cluster ZwCL~0926.5$+$30.26 \citep{eflu75,cor87,jkm+04,smh+17} with $B_\mathrm{eq}$ ranging from 0.89--1.6~\si{\micro\gauss}, which is consistent with \textcolor{black}{$B_\mathrm{eq}\approx2.7~\si{\micro\gauss}$} we find for NGC~1534.\par
With so few examples of of dead radio galaxies outside of rich clusters, adding counts to this population will only help in understanding the life-cycles of radio galaxies.

\subsection{The group environment}
NGC~1534 is catalogued as part of the HDC~269 and LDC~292 galaxy groups \citep[][]{chm+07}, which have line-of-sight velocity dispersions 119.3 and 198.7 \si{\kilo\m\per\s}. The high density group, HDC~269, has three members: NGC~1534 itself, NGC~1529, and 2MASX~J04111365-6242521; locations of these group members are indicated on Fig. \ref{fig:NGC1534_rgb} as cyan squares. The group velocity is calculated to be 5201 \si{\kilo\m\per\s} \citep{chm+07}. If NGC~1534 is indeed the host, then we see from the offset position of the optical galaxy and the thinnest point of the radio emission that NGC~1534 must have moved from its old lobes, leaving the radio plasma to gradually diffuse and lose energy to the intergalactic medium. The shortest timescale available for NGC~1534 for movement is if all velocity is in the transverse direction. If we assume that the transverse velocity is no more than its radial velocity, we can assume that the projected transverse velocity is given by the velocity dispersions which allows us to determine the maximum distance NGC~1534 could have drifted from it's original position or alternatively the minimum age of the emission had NCG~1534 drifted this far, assuming the greatest possible transverse velocity. Assuming the pinching point between the lobes was the original location of NGC~1534, we calculate a projected separation of 1.72~arcmin corresponding to 38.5~\si{\kilo\parsec}. With our assumed projected transverse velocities, this indicates a minimum age of the emission (and since NGC~1534 stopped producing it) of $190$~\si{\mega\year} (or $316$~\si{\mega\year}). Given that this is an underdense environment, the relic plasma will continue to move along the same path as NGC~1534, which pushes these times up further, \textcolor{black}{and movement in any direction not transverse will do the same}. This time scale is on the same order as the `off' spectral age of $\sim158$~Myr making it plausible that NGC1534 is the original host, despite the now imperfect alignment of the galaxy and the emission. \par

\subsection{\textcolor{black}{Implications for furture studies}}

It is worth noting that originally, the emission was not found in a survey but as part of a set of targeted observations. However, surveys such as GLEAM, the TFIR GMRT \footnote{Tata Institute of Fundamental Research Giant Metrewave Radio Telescope} Sky Survey \citep[TGSS alternate data release 1;][]{ijmf16}, the LOFAR Multifrequency Snapshot Sky Survey \citep[MSSS;][]{msss} and the LOFAR Two-metre Sky Survey \citep[LoTSS;][]{lotss1}, are providing the low frequency observations required to uncover a heretofore unseen population of faint, steep-spectrum sources. With GLEAM (and the MWA in general), the $u$--$v$ coverage offered by its short baseline observations allow for the detection of low-surface brightness, large-scale emission and where there exists overlap with the TGSS, there is the benefit of complementary, higher-resolution data to confirm, e.g., cores or other compact structure within the emission. \textcolor{black}{However, while emission with such large angular extent such as that presented here is comparatively rare, such low surface-brightness sources do not have to be so large. Smaller-scale, low surface-brightness emission (at higher redshift or otherwise) may be missed if using only the low-resolution GLEAM survey, though this may be alleviated somewhat with upcoming MWA Phase II surveys using the extended tile configuration, which will have a resolution on the order of two times that of GLEAM at a small cost to surface brightness sensitivity (Wayth et al., in preparation).}\par

\textcolor{black}{Remnant emission like that around NGC~1534 \citep[or `blob1';][]{bgm+16} would likely be missed in surveys focused within a small frequency band such as that offered by GLEAM due to their reasonably flat low-frequency spectra.} It has been suggested \citep[][but see also \citealt{skm03}]{mpm+11,bgm+16,bgm+17} that using the spectral curvature, $\mathrm{SPC}=\alpha_\mathrm{high}-\alpha_\mathrm{low}$, where $\alpha_\mathrm{high}$ and $\alpha_\mathrm{low}$ are high- and low-frequency spectral indices, respectively, would be a useful tool in detecting dead radio galaxies, as $\mathrm{SPC}<-0.5$ implies a non-active source. Such a diagnostic tool requires a good choice of $\alpha_\mathrm{high}$ and $\alpha_\mathrm{low}$, \textcolor{black}{though \citet{har17} note that even this may not be sufficient due to the different $\alpha_\mathrm{inj}$ between FR-I and FR-II sources. Furthermore, such a survey would require high- and low-frequency data }and at present the most sensitive, higher frequency southern sky counterpart to GLEAM is SUMSS, which may not be high enough in frequency. The upcoming Evolutionary Map of the Universe \citep[EMU;][]{emu1} with the Australia Square Kilometre Array Pathfinder \citep[ASKAP;][]{jbb+07} is expected to have a rms sensitivity on the order of 10~\si{\micro\jansky\per\beam}. This, coupled with its low-surface--brightness sensitivity and the frequency range 1130--1430~MHz, will give another high-frequency counterpart to GLEAM (and future MWA Phase II surveys) for searches of \textcolor{black}{remnant} radio galaxies, paving the way to vastly increase the detection rate of this population. \par
 
Within the context of searching for radio lobes of disk galaxies \citep[e.g.][]{sis+15}, the larger and possibly faint lobes of giant radio galaxies (e.g. \citealt{shsb05}), or searching for dead radio sources \citep[e.g.][]{mpm+11}, the low frequency surveys can be paired with optical surveys such as Pan-STARRS1 (Panoramic Survey Telescope And Rapid Response System; \citealt{kpc+10}, PS1; \citealt{tsl+12}, \citealt{cmm+16}) for $\delta > -30\degr$ or the SkyMapper Southern Sky Survey (SMSS; Wolf et al. in preparation \footnote{\url{http://skymapper.anu.edu.au/surveys/skymapper-southern-sky-survey/}}) for $\delta \lesssim +2\degr$ for confirmation of the optical host. On top of the additional depth of the surveys, PS1 has five optical bands from near infrared to blue and SMSS has six from NIR to NUV allowing better estimation of photometric redshifts which is vital when spectroscopic redshifts are unavailable.  Further in the future, southern spectroscopic surveys such as the Taipan Galaxy Survey \citep{taipan}, in conjunction with SMSS will provide an analogue to the Sloan Digital Sky Surveys \citep[SDSS;][specifically surveys such as the Legacy Survey; \citealt{sdssdr7}]{sdss1} and will facilitate surveys for disk galaxies hosting large-scale radio emission as well as surveys of dead or dying radio \textcolor{black}{sources.}

\section{Conclusion}
\textcolor{black}{
In this paper we have presented follow-up observations with the ATCA of the remnant emission surrounding the lenticular galaxy NGC~1534 originally detected by \citet{hje+15}. We combined this with new low frequency MWA data to study the emission from \textcolor{black}{72}~MHz to 19~GHz\textcolor{black}{, including polarimetric study in the ATCA 16~cm band.} We summarise the main results here.
\begin{itemize}
\item We find the northern lobe to be well-fit by a $\mathrm{CI}_{\mathrm{off}}$ model with a fixed $B_\mathrm{eq} \approx 2.7$~\si{\micro\gauss} and $\alpha_\mathrm{inj}=0.54\pm0.24$, obtain an estimate of the spectral age of the emission on the order of $\sim203$~Myr, having been active for only $\sim44$~Myr.
\item The ATCA 2.1- and 17-GHz data corroborate the notion that NGC~1534 has no significant core emission, with low-brightness, diffuse emission seen from the galaxy with no compact counterpart.
\item We find consistency between the radio and mid-infrared derived star-format rates for NGC~1534 with $S\!F\!R_{1.4} = 0.38 \pm 0.03$ M$_{\odot}$\,\si{\per\year} and $S\!F\!R_{22} = 0.37 \pm 0.04$ M$_{\odot}$\,\si{\per\year}, consistent with a lack of significant nuclear activity.
\item The northern lobe is shown to be highly linearly polarised at 1510 and 1942~MHz, with $m_\mathrm{P,1510}=42\pm13$~\% and $m_\mathrm{P,1942}=43\pm14$~\%. 
\item RM-synthesis of the region detects a significant Galactic foreground screen at $+33$~\si{\radian\per\m\squared} with an additional peak in the Faraday spectrum of the northern lobe at $-153$~\si{\radian\per\m\squared}.
\item Analysis of the position of NGC~1534 with the improved expected position of the host galaxy suggests that the host has drifted away from its original position over a time period commensurate with the cessation of AGN activities.  
\end{itemize}
From these findings, we see that the lenticular galaxy NGC~1534 and the surrouding emission is consistent with remnant radio galaxies, we confirm its classificaton as a rare `dead' radio galaxy not associated with a galaxy cluster. It additionally sits with the rare class of radio galaxies associated with dusty disk galaxies. With such a wealth of data soon available from upcoming and currently underway sky surveys at multiple wavelengths, we expect to be able to greatly expand both the number of such sources detected and using the panchromatic data available, better understand their host systems.}\par

\begin{acknowledgements}
SWD acknowledges a Doctoral Scholarship from Victoria University of Wellington and an Australian Government Research Training Programme scholarship administered through Curtin University. The Australia Telescope Compact Array is part of the Australia Telescope National Facility which is funded by the Australian Government for operation as a National Facility managed by CSIRO. \textcolor{black}{The authors would like to thank the anonymous referee for helpful comments and suggestions that helped to improve this paper.} \par
This research made use of \textsc{astropy}, a community-developed core \textsc{python} package for Astronomy \citep{astropy}, along with \textsc{aplpy}, an open-source plotting package for \textsc{python} hosted at \url{http://aplpy.github.com}. This research also made use of \textsc{NumPy} \citep{numpy}, \textsc{matplotlib} \citep{mpl}, and \textsc{iPython} \citep{ipython} which are part of the \textsc{SciPy} library for \textsc{python}: \url{https://www.scipy.org/}. This research has made use of the VizieR catalogue access tool, CDS, Strasbourg, France.  The original description of the VizieR service was described in \citet{vizier}. This research also made use of the NASA/IPAC Extragalactic Database (NED) which is operated by the Jet Propulsion Laboratory, California Institute of Technology, under contract with the National Aeronautics and Space Administration. The Digitized Sky Surveys were produced at the Space Telescope Science Institute under U.S. Government grant NAG W-2166. The images of these surveys are based on photographic data obtained using the Oschin Schmidt Telescope on Palomar Mountain and the UK Schmidt Telescope. The plates were processed into the present compressed digital form with the permission of these institutions.  
\end{acknowledgements}

\begin{appendix}

\end{appendix} 

\bibliographystyle{pasa-mnras}
\bibliography{bib_file,AGN_bib}

\end{document}